\documentclass[prb,aps,amsmath,amssymb,twocolumn]{revtex4}
\usepackage{graphicx}
\begin{document}
\title{Non-Born effects in scattering of electrons in a clean conducting tube}
\author{A.S. Ioselevich\footnote{e-mail: iossel@itp.ac.ru}}
\affiliation{Condensed-matter physics laboratory, National Research University Higher School of Economics, Moscow 101000, Russia,}
\affiliation{L. D. Landau Institute for Theoretical Physics, Moscow 119334, Russia}
\author{N.S.Pescherenko\footnote{e-mail: peshcherenko@phystech.edu}}
\affiliation{Moscow Institute of Physics and Technology, Moscow 141700, Russia}
\affiliation{Skolkovo Institute of Science and Technology, Moscow 121205, Russia}

\address{}
\date{\today}

\begin{abstract}
Quasi-one-dimensional systems demonstrate Van Hove singularities in the density of states $\nu_F$ and the resistivity $\rho$, occurring when the Fermi level $E$  crosses a bottom $E_N$ of some subband of transverse quantization.   We demonstrate that the character of smearing of the singularities  crucially depends on the concentration of impurities. There is a crossover concentration $n_c\propto |\lambda|$, $\lambda\ll 1$ being the dimensionless amplitude of scattering. For $n\gg n_c$  the singularities are simply rounded at $\varepsilon\equiv E-E_N\sim \tau^{-1}$ -- the  Born scattering rate.
For $n\ll n_c$ the non-Born effects in scattering become essential despite $\lambda\ll 1$. The peak of the resistivity is asymmetrically split in a Fano-resonance manner (however with a more complex structure). Namely, for $\varepsilon>0$ there is a broad maximum at $\varepsilon\propto \lambda^2$ while for $\varepsilon<0$ there is a deep minimum at $|\varepsilon|\propto n^2\ll \lambda^2$. The behaviour of $\rho$  below the minimum depends on the sign of $\lambda$. In case of repulsion $\rho$ monotonically grows with $|\varepsilon|$ and saturates for $|\varepsilon|\gg \lambda^2$. In case of attraction $\rho$ has sharp maximum at $|\varepsilon|\propto \lambda^2$. The latter feature is due to resonant scattering by quasistationary bound states that inevitably arise just below the bottom of each subband for any attracting impurity.
\end{abstract}

\maketitle

\section{Introduction\label{Introduction}}

In this study we consider clean multichannel quasi-one-dimensional metallic systems: wires, tubes, strips etc. We revisit a seemingly well-understood problem of semiclassical (i.e. without localization effects) resistivity for such systems in the presence of weak short-range impurities at low concentration. It is well known, that  this resistivity  (as well as the density of states at the Fermi level) has a square-root Van Hove singularities as a function of the Fermi level position $E$,  occurring when $E$ crosses a bottom $E_N$ of certain  subband of transverse quantization \cite{VanHove}. These singularities are expected to be smeared due to scattering of electrons by impurities and (at least in the Born approximation) the width  of the peak  $\Gamma_{\rm B}\sim\tau_{\min}^{-1}$ can be estimated as an electronic scattering rate at maximum of resistivity. This smearing was theoretically studied within the self-consistent Born approximation by different groups of authors \cite{self-consistent, self-consistent1, self-consistent2, Lee, Peeters, HuegleEgger2002}.

We demonstrate that the above picture is valid only if the concentration of impurities is relatively high while for low concentration  due to specifics of the quasi-one-dimensional systems  the non-Born effects  become essential despite the nominal weakness of scattering. These effects lead to dramatic restructuring of the Van Hove singularities. 

Complex asymmetric features were experimentally observed in many quasi-one-dimensional systems, such as nanotubes (both single-wall \cite{single-wall0, single-wall} and multi-wall \cite{multi-wall1,multi-wall2} ones). These features were attributed to Fano resonance \cite{fano}, arising due to interference of the scattering at some narrow resonant state with the scattering at background continuum. The $E$-dependence of resistivity $\rho$  at the Fano resonance is usually described by the formula
\begin{align}
\rho(E)\propto\frac{(E-E_N+q\Gamma/2)^2}{(E-E_N)^2+(\Gamma/2)^2}
	\label{fano}
\end{align}
with phenomenological  parameters $q$ and $\Gamma$ (see, e.g., \cite{fano1}).
There were attempts \cite{single-wall0, single-wall, multi-wall1} to fit the experimental data on the Van Hove singularities in nanotubes with the formula \eqref{fano} with an appropriate choice of $\Gamma$ and $q$. We will show, however, that this phenomenological expression is not sufficient to describe the entire zoo of possible $\rho(E)$ shapes. In this paper we will give a microscopic derivation of the actual $\rho(E)$ dependence. The main ingredient of our theory is the non-Born effects in scattering.

\subsection{Statement of the problem\label{Statement of the problem}}

In this paper we restrict our consideration to only one simple realization of quasi-one-dimensional system: a single-wall metallic tube in a strong longitudinal magnetic field.  The zero (or weak) field case is more difficult for theoretical study because of the chiral degeneracy of the electronic states that is only lifted due to an interaction with an impurity. Other quasi-one-dimensional variants such as a conducting strip involve some additional complications  due to nonequivalence of positions of different impurities with respect to the nodes of the wave function. All these effects are quite interesting and they will be discussed elsewhere. 

Besides the simplicity of the theoretical interpretation the case of strong magnetic field is convenient practically since the changing of magnetic field is an effective instrument for tuning the distance $E-E_N$, so it is easy to sweep the Van Hove singularity in a controllable way. 

Oscillations of the longitudinal resistivity with the magnetic flux $\Phi$ threading the tube is a well known effect that was experimentally observed in various tubes and wires (especially semimetallic) \cite{Brandt77, Brandt82, Nikolaeva2008}. These oscillations are the direct manifestation of the Aharonov-Bohm effect \cite{AB59} -- the interference of electronic waves with opposite chiralities. From the semiclassical point of view it is instructive to write the resistivity $\rho$ in the form of Fourier series:
\begin{align}
\rho=\rho_0\left(1+\sum_{n=1}^\infty A_n\cos(\pi n\Phi/\Phi_0)\right)
	\label{gen-rho}
\end{align}
where $\Phi_0=\pi c\hbar/e=ch/2e$ is the flux quantum. The oscillations can be observed in both dirty and clean systems. As it was shown in a seminal paper by Altshuler, Aronov and Spivak \cite{AAS81} in dirty (diffusive) systems the odd-$n$ harmonics of the Aharonov-Bohm oscillations \eqref{gen-rho} are washed out due to strong variations in the length of different diffusive trajectories that lead to randomisation of the non-magnetic part of the phases of electronic wave-functions. The even-$n$ harmonics -- the oscillations associated with a special sort of trajectories (the ones containing  closed topologically non-trivial loops on the cylinder) survive the randomisation. This effect was observed in experiments (see \cite{AASSS82,AS87}). The odd-$n$ harmonics are in general very fragile: they may be suppressed also in nominally clean systems \cite{Nikolaeva2008} due to the fluctuations of the tube's parameters: e.g., the radius \cite{Ioselevich2015}. The even-$n$ harmonics are less fragile but still, in the presence of any kind of disorder the amplitudes $A_n$ rapidly decrease with $n$ so that the oscillations in the imperfect systems usually look roughly harmonic.

\begin{figure}[ht]
\includegraphics[width=0.9\linewidth]{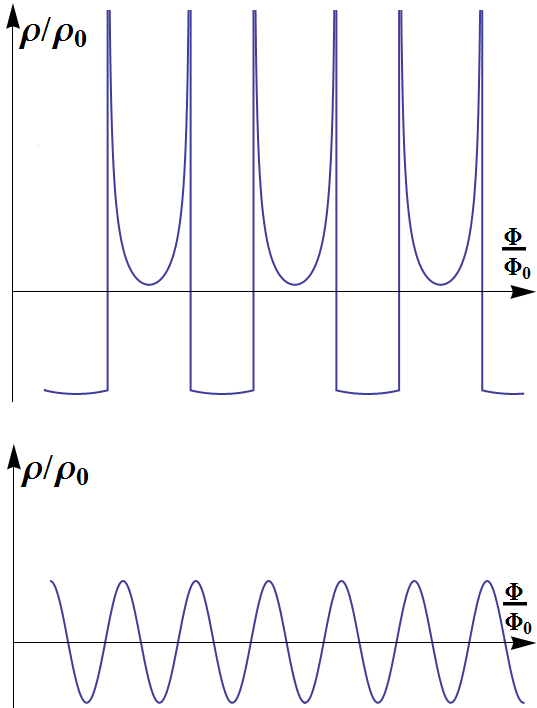}
\caption{$\rho(\Phi)$ dependence for clean and dirty cases. Top: clean case, $\rho(\Phi)$ is periodic with a period $2\Phi_{0}$ and Van Hove square root singularities are present for $\Phi=2n\Phi_0$. Bottom: dirty case, $\rho(\Phi)$ is periodic with a period $\Phi_{0}$ - odd harmonics are suppressed.}
\label{flux_dependence}
\end{figure}

It is not the case for the geometrically perfect clean systems where $A_n$ decrease with $n$ only as $n^{-1/2}$ so that the series diverges at $\Phi\to 2M\Phi_0$ with integer $M$. This divergency is nothing else but the Van Hove singularity (see, e.g., \cite{Ioselevich2015}). So, the shape of oscillations is very different for perfect and for imperfect systems (see Fig.\ref{flux_dependence}).

It this work we concentrate on geometrically perfect tubes with low concentration of weak short-range impurities, where one can expect strongly unharmonic oscillations dominated by the Van Hove singularities as in the upper panel of Fig.\ref{flux_dependence}.

Thus, we consider a single-wall tube of radius $R$ threaded by magnetic flux $\Phi$. The tube is  supposed to be cut from a sheet of two-dimensional metal with simple quadratic spectrum \cite{spectrum} of electrons $E=\hbar^2{\bf k}^2/2m^*$.  Impurities are embedded in this sheet with two-dimensional concentration $n_2$. They are supposed to be short-range and weakly scattering ones.

\subsection{Principal approximations\label{The principal approximations}}

It is convenient to measure all the energies in the units of $E_0=\hbar^2/2m^*R^2$ and we will assume the semiclassical condition throughout this paper:
\begin{align}
  \varepsilon_0\gg1,\quad  	\varepsilon_0\equiv E/E_0, \quad N\sim\varepsilon_0^{1/2}\gg1.\label{semic1}
\end{align}
where $N$ is the label of a subband whose bottom is the closest to the Fermi level and has the meaning of the number of open channels in the system.

The magnetic field is assumed to be strong enough so that the splitting between $E_N$ and $E_{-N}$ is larger than the width of the peaks $\Gamma$. Besides that, the parameter $2\sqrt{\varepsilon_0}$ should not be close  to any  integer $K$ to avoid resonance between the subbands with $m=N$ and $m'=N\pm 2\sqrt{\varepsilon_0}$.

 All the interesting effects associated with the Van Hove singularities occur in the range where 
\begin{align}
  |\varepsilon|\ll 1,\qquad  \varepsilon \equiv (E-E_N)/ E_0. \label{semic001}
\end{align}

There are two important dimensionless small parameters in this problem:

(i) The dimensionless concentration of impurities
\begin{align}
    	n\equiv n_2 (2\pi R)^2, \qquad n\ll 1.\label{semic2}
\end{align}
 It is assumed to be small which in particular means that the average distance between impurities is larger than the transverse size of the system.
 
(ii) Dimensionless scattering amplitude
\begin{align}
 \Lambda_{2d}=\lambda-i\lambda^2 \label{semic3f}
\end{align}
of the background two-dimensional problem ($\lambda>0$ corresponds to repulsion, $\lambda<0$ -- to attraction). It is also supposed to be small:
\begin{align}
    	|\lambda|\ll 1. \label{semic3}
\end{align}
The imaginary part of complex $\Lambda_{2d}$ in \eqref{semic3f} is necessary to fulfil   the unitarity requirement \cite{opticaltheorem} (the optical theorem):
\begin{align}
{\rm Im}\;\Lambda_{2d}=-|\Lambda_{2d}|^2.
	\label{unitarity10za}
\end{align}
Actually we will need this imaginary part only for proper treatment of quasistationary states arising in the case of attracting $\lambda<0$. In all other cases we can simply put $\Lambda_{2d}\to\lambda$.

(iii) The third condition is imposed on the length $L$ of the tube: it should satisfy the inequality
\begin{align}
l(\varepsilon)\ll L\ll {\cal L}_{\rm loc}(\varepsilon),
	\label{unitarity10zatr}
\end{align}
where $l(\varepsilon)$ is the mean free path and ${\cal L}_{\rm loc}\sim Nl(\varepsilon)$ is the localization length. The large parameter $N\gg 1$ assures at least the possibility for inequality \eqref{unitarity10zatr} to be fulfilled.

Indeed, it is very well known that weak localization effects in quasi-one-dimensional systems lead (in the absence of inelastic processes) to an ultimate localization on all electronic states. However, for the tubes with lengths $L$ in the range \eqref{unitarity10zatr} the localization corrections are still small so that the results obtained throughout this paper are well justified and should give a valid expressions for the resistivity $\rho(\varepsilon)$. Moreover, these results provide a possibility to estimate the dependence of the localization length on the parameter $\varepsilon$:
\begin{align}
{\cal L}_{\rm loc}(\varepsilon)=\left[e^2\rho(\varepsilon)\right]^{-1}.
	\label{unitarity10zat}
\end{align}

\subsection{The structure of the paper\label{The structure of the paper}}

The structure of the paper is as follows:
In Section \ref{The principal results} we bring together all the principal results of the paper. In Section \ref{Ideal system} we briefly remind the well-known facts about quantum mechanics of an electron on  a tube threaded by magnetic field. In Section \ref{Born scattering by short-range impurities} we discuss the  scattering of electrons near the Van Hove singularity within the Born approximation. In Section \ref{Beyond the Born approximation} we discuss the non-Born effects for scattering of electrons in the cylinder geometry and derive the corresponding renormalization of the scattering amplitude. In the case of attracting potential we find the poles in the scattering matrix that are related to quasistationary states under the bottom of each subband. In Section \ref{Non-Born effects in resistivity} we consider the manifestations of the non-Born effects in resistivity in the single-impurity approximation. In particular we demonstrate that in this approximation the resistivity vanishes exactly at the Van Hove singularity. In Section \ref{Finite concentration of impurities} we estimate the effects of interference between scattering events at different impurities, resolve the zero-resistivity paradox of the preceeding section and estimate the minimal resistivity. In Section \ref{Smearing of the resonant peak} we discuss the inhomogeneous broadening of the peaks in the resistivity that arise due to resonant scattering at quasistationqry states. In Section \ref{Systems with different sorts of impurities} we explore the effects that should arise if impurity with different effective scattering amplitudes are present in the system. In Section \ref{discussion and conclusion} we summarize the results and outline the directions of future research. In the Appendices \ref{self-consistent Born approximation} and \ref{self-consistent non-Born approximation} we evaluate the behaviour of the system in the immediate vicinity of the Van Hove singularity (where the single-impurity approximation breaks down) using the self-consistent  approximation and explain the effect of a ``catastrophic drop'' (almost a jump!) of resistivity just below the smeared singularity.

\section{The principal results\label{The principal results}}

The number of physical scenarios and distinct ranges of parameters considered in this paper is large. Therefore we find it reasonable to start with the list of different regimes and principal results.

\subsection{The Born approximation\label{The Born approximation0}}

 Away from the Van Hove singularities (at $|\varepsilon|\gg 1$) the applicability of the  Born approximation requires only the condition \eqref{semic3}. Here the system behaves simply as a classic piece of the background two-dimensional material. The density of states, the resistivity and the scattering rate (the latter is being measured in units of $E_0$) are
 \begin{align}
 \nu_0=m^*R,\quad   	\rho_{0}= \frac{1}{e^2\varepsilon_0}\frac{1}{\tau_0},\quad \frac{1}{\tau_0}=2n\left(\frac{\lambda}{\pi}\right)^{2}. \label{semic3a}
\end{align}

In all cases the main contribution to the current comes from the one-dimensional subbands with labels $m$ that are not very close to $N$ because for $m\approx N$ the longitudinal velocity of electrons with energy $E$ tends to zero. However, the role of the $N$-band becomes very important near the singularity when $E\to E_N$. Indeed, when the total density of states
\begin{align}
    	\nu(\varepsilon)=\nu_0\left(1+\frac{\theta(\varepsilon)}{\pi\sqrt{\varepsilon}}\right), \label{semic3afre}
\end{align}
is dominated by the second term (the contribution of the resonant $N$-band), the electrons from the current-carrying bands (those with labels $m\sim N/2$) are scattered predominantly to the resonant one (with $m=N$). Near the singularity the properties of electrons from the resonant band differ from the properties of all others. For a general quasi-one-dimensional system there are in principle two distinct scattering amplitudes and corresponding rates: $\lambda_1$ and $\tau^{-1}_1(\varepsilon)$ describe the scattering from the current-carrying bands to the resonant one while $\lambda_2$ and $\tau_2^{-1}(\varepsilon)$ correspond to scattering within the resonant band. The rate $\tau^{-1}_1(\varepsilon)$ directly determines the mean free path and the resistivity
\begin{align}
    	\rho(\varepsilon)= \frac{1}{e^2\varepsilon_0}\frac{1}{\tau_1(\varepsilon)}=\frac{1}{e^2Nl(\varepsilon)}, \quad l(\varepsilon)=N\tau_1(\varepsilon),
	\label{semic3afr}
\end{align}
where we have used the obvious relations $l\sim v_F\tau$ and $v_F\sim E^{1/2}\sim N$.
The rate $\tau_2^{-1}(\varepsilon)$ is responsible for smearing of the singularity in the density of states and is relevant only in the immediate vicinity of the singularity. However, we show in Section \ref{Born scattering by short-range impurities} that for the case of a tube
\begin{align}
    	\tau_2(\varepsilon)=\tau_1(\varepsilon)\equiv \tau. \label{tau1tau2}
\end{align}

 We will show, that close to the singularity the Born approximation remains valid only if the dimensionless concentration $n$ of impurities is relatively high. 
 
 Let us first assume that this condition is fulfilled and estimate the width of the smeared Van Hove singularity. The scattering rate is proportional to the density of final states so that
 \begin{align}
    	 \frac{1}{\tau_1(\varepsilon)}=\frac{1}{\tau_{0}}\frac{\nu(\varepsilon)}{\nu_0}. \label{semic3aa}
\end{align}
The width $\Gamma_{\rm B}$ of the peak in the density of states (and in the resistivity at the same time) may be estimated from the condition 
 \begin{align}
\tau^{-1}_1\left(\varepsilon\sim\Gamma_{\rm B}\right)\sim \Gamma_{\rm B}    . \label{semic3ab}
\end{align}
As a result, the Van Hove singularity is smeared on the scale $|\varepsilon|\sim \Gamma_{\rm B}$
\begin{align}
    	\Gamma_{\rm B}\sim \left(\frac{n}{\pi}\right)^{2/3}\left(\frac{\lambda}{\pi}\right)^{4/3}\gg\frac{1}{\tau_0},\label{semic5m}\\ \rho_{\rm B}^{\max}\sim \frac{1}{e^2\varepsilon_0}\left(\frac{n}{\pi}\right)^{2/3}\left(\frac{\lambda}{\pi}\right)^{4/3}\gg \rho_{0}. \label{semic5}
\end{align}

\subsection{The origin of non-Born effects\label{Non-Born effects0}}

The origin of the special importance of non-Born effects in quasi-one-dimensional systems is  renormalization of the scattering matrix that is dramatically enhanced near a Van Hove singularity. In the case of tube this matrix can be effectively reduced to a single complex constant  $\Lambda(\varepsilon)$ that can be found from the Dyson equation. As a result
\begin{align}
\lambda\to\Lambda(\varepsilon)\approx\lambda\left\{1-\frac{\Lambda_{2d}}{\pi\sqrt{\varepsilon}}\right\}^{-1}
\label{semic008}
\end{align}
 From \eqref{semic008} it is clear that the energy scale
\begin{align}
    \varepsilon_{\rm nB}= \left(\lambda/\pi\right)^{2}\ll 1,\label{semic5o}
\end{align}
measures the range near the singularity where the non-Born effects are considerable. In particular, we see that if, due to low concentration of impurities, the Born scattering rate is low enough: 
\begin{align}
    \Gamma_{\rm B}<\varepsilon_{\rm nB},\label{semic5ll}
\end{align}
then the non-Born effcts  have chance to come into play in the range $\Gamma<|\varepsilon|<\varepsilon_{\rm nB}$. Substituting the explicit formulas \eqref{semic5m} and \eqref{semic5o} to the condition \eqref{semic5ll}, we arrive at the criterion 
\begin{align}
   n<n_c,\qquad 	n_c=|\lambda|/\pi. \label{semic4d}
\end{align}
of the breakdown of the Born approximation in the vicinity of the singularity. Under the opposite condition the Born approximation is sufficient for all $\varepsilon$.

It is convenient to rewrite \eqref{semic008} in the form
\begin{align}
\lambda\to\Lambda(\epsilon)\approx\frac{\lambda}{1-[{\rm sign}(\lambda)-i|\lambda|](-\epsilon)^{-1/2}},\quad \epsilon\equiv \frac{\varepsilon}{\varepsilon_{\rm nB}}.
\label{semic008aq}
\end{align}

\subsection{The  non-Born effects in resistivity: repulsing impurities\label{The  non-Born effects in resistivity: repulsing impurities0}}

At low concentration of impurities $n\ll n_c$ the shape of the $\rho(\epsilon)$ dependence in the vicinity of Van Hove singularities is strongly modified by non-Born effects in scattering.

\begin{figure}[ht]
\includegraphics[width=0.9\linewidth]{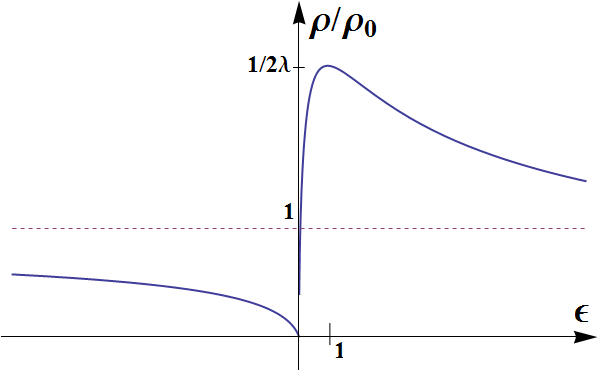}
\caption{Dependence of the resistivity on the position of the Fermi level for repulsing impurities in the case of strong non-Born regime low concentration of impurities $n\ll n_c$. Note that $\rho(\epsilon)$ vanishes as $\epsilon\to 0$: it is an artefact of the single-impurity approximation that is not applicable at $\epsilon\lesssim \epsilon_{\min}\ll 1$. The horizontal asymptote (dashed line) corresponds to $\rho=\rho_0$.}
\label{resist-repuls}
\end{figure}

A narrow peak at  $\epsilon=0$ is replaced by a broad one slightly above the bottom -- with the maximum at $\epsilon\sim 1$ and the width $\Gamma^{(+)}_{\rm nB}\sim \varepsilon_{\rm nB}$, independent of the concentration $n$. 
The shape of this broad peak can be found explicitly:
\begin{align}
\frac{1}{\tau(\varepsilon)}=2\left(\frac{n}{\pi}\right)\left(\frac{\lambda}{\pi}\right)F(\epsilon), \quad \epsilon\equiv \varepsilon/\varepsilon_{\rm nB},\label{semi65a}
\end{align}
\begin{align}
    	 F(\epsilon)=
(\epsilon^{1/2}+\epsilon^{-1/2})^{-1}
\end{align}
The maximal (in the range $\varepsilon>0$) resistivity 
\begin{align}
    	 \rho_{\rm nB}^{\max(+)}\sim \frac{2}{e^2\varepsilon_0}\left(\frac{n}{\pi}\right)\left(\frac{\lambda}{\pi}\right)F_{\max}\ll \rho_{\rm B}^{\max}, \label{semi65}
\end{align}
is reached at $\epsilon=\epsilon_{\max}$,
where
\begin{align}
    	 F_{\max}=1/2,	\quad
 \epsilon_{\max}=1\label{semi8}
\end{align}
The function $F(\epsilon)$ is shown in Fig. \ref{resist-repuls}. At $\epsilon\gg 1$ it has asymptotics $F(\epsilon)\approx \epsilon^{-1/2}$ that corresponds to the standard Van Hove singularity. The height of the broad peak is much less than it would be within the Born approximation but still is much higher than the background resistivity $\rho_0$.

The behaviour of the resistivity above the Van Hove singularity (for $\varepsilon>0$), described by \eqref{semi65a}, does not depend on the sign of $\lambda$, it is the same for attracting and repulsing impurities. It is not the case for the range $\varepsilon<0$ below the singularity. For repulsing impurities we obtain
\begin{align}
    	 \frac{1}{\tau(\varepsilon)}=2\pi \left(\frac{n}{\pi}\right)\left(\frac{\lambda}{\pi}\right)^2\tilde{F}(\epsilon), \label{semi65ae}
\end{align}
\begin{align}
    	 \tilde{F}(\epsilon)=
[1+(-\epsilon)^{-1/2}]^{-2},
\label{semi65be}
\end{align}
so that $\rho(\varepsilon)$ monotonically increases with $|\varepsilon|$ and saturates at $\rho=\rho_0$.

It is easy to see that, as it formally follows from \eqref{semi65be}, the resistivity $\rho(\varepsilon)$ should vanish for $\varepsilon\to 0$ from either side. Indeed, for $|\epsilon|\ll 1$
\begin{align}
 F(\epsilon)\approx
\epsilon^{1/2} \quad 	 \tilde{F}(\epsilon)\approx
-\epsilon.
\label{semi65be2}
\end{align}

 Of course we immediately suspect that in reality the decrease of resistivity will be ultimately stopped by some additional effect (and this is indeed so, see below). But anyway, a dramatic suppression of resistivity in the narrow vicinity of the Van Hove point is an important phenomenon. Physically it is a result of destructive interference of partial electronic waves with different winding numbers.

\subsection{Attracting impurities, quasistationary states and resonant scattering\label{Attracting impurities, quasistationary states and resonant scattering}}

As we have already mentioned in previous section, the behaviour of resistivity above the singularity is identical for repulsing and attracting impurities. However, below the singularity the attracting impurities introduce some nice additional physics. 

It can be shown that, besides the true bound state with the energy below the bottom of the lowest subband of the electronic spectrum of the cylinder, a weakly attracting short-range impurity produces an infinite series of quasistationary states: one such state below the bottom of each band. In this paper we concentrate on the quasistationary states associated with the quasiclassic subbands (those, with large $N\gg 1$). In particular we show that for $\lambda<0$ the scattering amplitude \eqref{semic008aq} has a pole at $\epsilon=-1+2i|\lambda|$ (or at $\varepsilon=(-1+2i|\lambda|)\varepsilon_{\rm nB}$ in other notation). This pole corresponds to a quasistationary state with a relatively small  decay rate. In the case of cylinder these poles are identical for all impurities and, since electrons can be scattered by these resonances, the latter lead to formation of sharp maxima in resistivity for $\varepsilon<0$ and $\lambda<0$:
\begin{align}
F(\epsilon)\approx\left\{
	\begin{aligned}
\frac{1}{(1-|\epsilon|^{-1/2})^2},\quad	&\mbox{for  $|1-|\epsilon||\gg |\lambda|$},\\
\frac{4}{(1-|\epsilon|)^2+4\lambda^2},\quad	&\mbox{for  $|1-|\epsilon||\lesssim |\lambda|$},
	\end{aligned}\right.
\label{self-consist2ssk0}		
\end{align}
This result is illustrated by Fig. \ref{resist-attract}:
\begin{figure}[ht]
\includegraphics[width=0.9\linewidth]{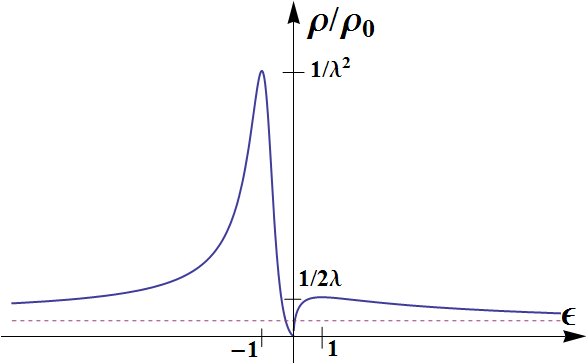}
\caption{The same as in Fig. \ref{resist-repuls} but for attracting impurities. The sharp maximum at $\epsilon<0$ arises due to resonant scattering at quasistationary states.}
\label{resist-attract}
\end{figure}

The maximal (in the range $\epsilon<0$) resistivity is reached at $\epsilon=-1$,
\begin{align}
    	 \rho_{\rm nB}^{\max(-)}\sim \frac{1}{e^2\varepsilon_0}\frac{2n}{\pi^2}, \label{semi65u}
\end{align}
the width of this maximum is $\Gamma_{\rm nB}^{(-)}=4|\lambda|\varepsilon_{\rm nB}$.

The physical origin of the quasistationary states that exist slightly below each of the subbands is as follows. Semiclassical trajectories of electrons with energies near the bottom of subband are almost closed; if an electron with such energy has passed near certain impurity once then it will do so again, and many times. Therefore the attraction to impurity is strongly enhanced and the bound state is formed. An alternative way of thinking is just to neglect in the leading approximation all the transitions from the resonant band to all others. The arising strictly one-dimensional problem grants a bound state for arbitrary weak attraction. Taking the transitions to nonresonant bands into account perturbatively leads to the finite decay rate of the state.

\subsection{The minimum of resistivity\label{The minimum of resistivity}}

All the effects described above are the single impurity ones. Their origin is the coherent multiple scattering by the same impurity which fact is manifested in the linear dependence of resistivity on the concentration $n$. To reveal the mechanism that limits the suppression of resistivity at $\epsilon\to0$ and to estimate the resistivity at its minimum one has to find the scattering rate $\tau^{-1}(\varepsilon)$ from \eqref{semi65ae} in the range $|\varepsilon|\ll \varepsilon_{\rm nB}$:
 \begin{align}
\frac{\tau_0}{\tau(\varepsilon)}=|\varepsilon|\left(1+\frac{\theta(\varepsilon)}{|\lambda|\sqrt{\varepsilon}}\right)
\label{selfuuw}		
\end{align}
The width $\Gamma_{\rm nB}$ of the peak in the density of states can be estimated from the condition
 \begin{align}
\Gamma_{\rm nB}\sim\tau^{-1}\left(\varepsilon\sim +\Gamma_{\rm nB}\right),\label{selfuuw0}	
\end{align}
and we get 
 \begin{align}
\Gamma_{\rm nB}\sim \varepsilon_{\min}\equiv (n/\pi)^2 \ll \varepsilon_{\rm nB}.
\label{selfuuw1}		
\end{align}
Note that this width does not depend on $\lambda$. At $\varepsilon<0$ the resonant contribution to the density of states rapidly drops on the same energy scale so that the factor $\nu(\varepsilon)$ becomes of order of $\nu_0$ already at $\varepsilon\sim -\varepsilon_{\min}$. As a result, the resistivity  has a minimum at $\varepsilon=\varepsilon_{\rm dip}$, where 
 \begin{align}
\varepsilon_{\rm dip}<0,\quad |\varepsilon_{\rm dip}|\sim\varepsilon_{\min}. \label{selfuuw2}		
\end{align}
The scattering rate and the resistivity at minimum are
\begin{align}
\frac{1}{\tau_{\rm dip}}\sim n^3,\quad
\rho_{\rm dip}\sim\frac{n^3}{e^2\varepsilon_0},
\label{deni2pp4puiew33}
\end{align}
and do not depend on the scattering amplitude $\lambda$. Thus, there is a deep and narrow universal minimum of resistivity slightly below the bare Van Hove singularity, the resistivity in the minimum depends on $n$ superlinearly.

\section{An ideal tube\label{Ideal system}}

We consider a tube of radius $R$  threaded by a magnetic flux $\Phi=\pi R^2H$ (the magnetic field $H$ is oriented along the axis of a cylinder $z$). 

\begin{figure}[ht]
\includegraphics[width=0.9\linewidth]{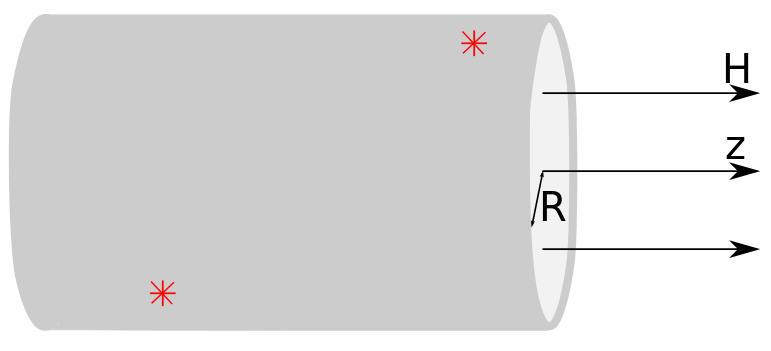}
\caption{Thin conducting tube, threaded by magnetic field $H$. Impurities (shown as stars) are embedded in the tube. Electrons live on the surface of the cylinder.}
\label{setup}
\end{figure}

 Electrons in the tube have the following spectrum and wave functions:
\begin{align}
    	\psi_{mk}(\phi,z)=(2\pi)^{-1/2}\exp\{ikz+im\phi\},\\
	 E_{mk}=\frac{\hbar^2k^2}{2m^*}+E_m, \label{main-cond2}
\end{align}
\begin{align}
E_m=E_0(m+\Phi/2\Phi_0)^2,\quad E_0=\frac{\hbar^2}{2m^*R^2}
	\label{main-cond2a}
\end{align}
where $m\in Z$ is the azymuthal quantum number, $k$ is the momentum along the cylinders axis and $\Phi_0=\pi c\hbar/e=ch/2e$ is the flux quantum. $E_m$ has the meaning of position of the bottom of $m$-th one-dimensional subband. Actually we have introduced the magnetic field as a tool of easy shifting of the Fermi level in the system but all the physics described below is present already in the case $H=0$.

  The density of states in each subband
\begin{align}
   \nu_m(E)=\int \frac{dk}{2\pi}
   \delta\left(E-E_m-\frac{k^2}{2m^*}\right)=\nonumber\\=\frac{2}{2\pi}\sqrt{\frac{m^*}{2(E-E_m)}}\theta(E-E_m),
   \label{main-cond2rs}
\end{align}
The factor 2 arises because the equation  $E-E_m-\frac{k^2}{2m^*}=0$ has two roots $k=\pm\sqrt{2m^*(E-E_m)}$.

\begin{figure}[ht]
\includegraphics[width=0.9\linewidth]{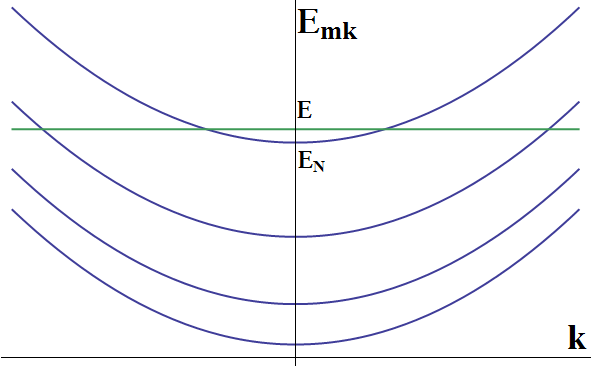}
\caption{Spectrum of an electron on a surface of an ideal cylinder. Subbands of the transverse quantization are shown. The Fermi level $E$ crosses all the subbands with $m\leq N$.}
\label{subbands}
\end{figure}

The total density of states
\begin{align}
   \nu(E)=\sum_m\nu_m(E)=-\frac{1}{\pi}{\rm Im}\,g(E),\\
   g(E)\equiv G_E^{(0)}(0,0)=\sum_m\int\frac{dk}{2\pi}\frac{1}{E-E_{km}+i0}=\nonumber\\=\sum_m\sqrt{\frac{m^*}{2(E_m-E)}},	
   \label{deni}
\end{align}
$G_E^{(0)}(0,0)$ being the one-point retarded Green function of an ideal wire. Strictly speaking, the real part of $g$ diverges. The recipe how to deal with this divergency will be discussed somewhat later. Now we just mention that the divergent part is energy-independent and therefore can be removed by a constant shift of the energy.

In the main part of this paper we will measure all energies in the units of $E_0$ and all distances in the units of $2\pi R$:
\begin{align}
E\equiv E_0\varepsilon_0,\quad  E-E_m\equiv E_0\varepsilon_m,\quad \nu_m(E)\equiv\frac{\nu_m(\varepsilon)}{2\pi RE_0},  \label{units0}
\end{align}
\begin{align}
 \nu_m(\varepsilon)=\frac{1}{\sqrt{\varepsilon_m}}\theta(\varepsilon_m),\quad g(\varepsilon)=\sum_m\frac{\pi}{\sqrt{-\varepsilon_m}}.
   \label{units}
\end{align}

We are interested in semiclassical case when $E_0\ll E$ or $\varepsilon\gg 1$. Then, in the leading semiclassical approximation
\begin{align}
   \nu(\varepsilon)=\sum_{m=-\infty}^\infty\nu_m(\varepsilon)\approx\nu_0=\int_0^{\varepsilon_0}\frac{d\varepsilon_{m}}{\sqrt{\varepsilon_m(\varepsilon_0-\varepsilon_m)}}=\pi.
   \label{deni1}
\end{align}
This result is valid for all $\varepsilon$ except narrow intervals in the vicinity of points where $\varepsilon_m=0$ for some $m$.

The condition of strong magnetic field reads
\begin{align}
\varepsilon_N-\varepsilon_{-N}\sim \sqrt{\varepsilon_0}\Phi/\Phi_0\gg\Gamma, 
\end{align}
where $\Gamma$ is the broadening of peaks and $N=\sqrt{\varepsilon_0}$ denotes the closest to Fermi level $E$ subband.

In the entire range of variation of $\epsilon$ one can write
\begin{align}
   \nu^{(0)}(\varepsilon)\approx\nu_0
		  \left(1+\frac{\theta(\varepsilon)}{\pi\sqrt{\varepsilon}}\right), 
   \label{deniww}
\end{align}
where we have introduced $\varepsilon\equiv\varepsilon_N$ for brevity.

Under the semiclassical condition $N\gg 1$ the result \eqref{deni1} is not valid in the vicinity of the Van Hove singularity for $\varepsilon\lesssim 1$ where the second -- resonant -- term in \eqref{deniww}  is anomalously large.
We see that for
\begin{align}
\varepsilon >0,\quad \varepsilon\ll 1   \label{deni2}
\end{align}
the inequality $\nu_N(\varepsilon)\gg \nu_0$ holds: the density of states is indeed dominated by the second term in \eqref{deniww} -- the contribution of the $N$-subband.
Note that in the semiclassical limit $N\gg 1$ the different peaks in the function $\nu(\varepsilon)$ are strictly identical.
 
 \section{Born scattering by short-range impurities\label{Born scattering by short-range impurities}}
 
Our first step is finding the longitudinal resistivity of the tube using the Drude and Born approximations. We consider weak short range impurities with the hamiltonian
\begin{align}
	\hat{H}=\hat{H}_0+V\delta({\bf r}-{\bf r}_0),\qquad \hat{H}_0=-\nabla^2/2m^*
	\label{imp-potential}
\end{align}
where $\delta({\bf r})\equiv \frac{1}{R}\delta(z-z_0)\delta(\phi-\phi_0)$ is a two-dimensional delta-function and ${\bf r}_0$ denotes the position of the impurity on the wall of the tube. Let us find the self-energy for an electron
\begin{align}
\Sigma_{km}=\frac{2\pi Rn_{\rm imp}^{(2)}}{E_0}\sum_{m'}\int\frac{dk'}{2\pi}|V_{kk'mm'}|^2G^{(0)}_{k'm'}
	\label{self-energy}
\end{align}
Since for the short range potential \eqref{imp-potential} 
\begin{align}
V_{kk'mm'}=\frac{V}{2\pi R}\exp\{i(m-m')\phi_0+i(k-k')z_0\},	\label{self-energyqsw}
\end{align}
$|V_{kk'mm'}|^2\equiv |V|^2$ depends neither on $km$, nor on $k'm'$, we conclude that $\Sigma_{km}=\Sigma(E_{km})$ is a function of only the total energy $E$. In our dimensionless variables we get:
\begin{align}
\Sigma(\varepsilon)=\frac{g(\varepsilon)}{2\pi\nu_0\tau_0},
	\label{self-energy1}
\end{align}
where $\tau_0$ is the dimensionless scattering time for an electron away from the resonance (i.e., for $\varepsilon\gg 1$):
\begin{align}
	\tau_0^{-1}=\frac{m^* V^2n_2}{E_0}=2n(\lambda/\pi)^2,\label{life-timeb1ep}
\end{align}
 the dimensionless Born scattering amplitude
 \begin{align}
	\lambda= 	m^*V/2,\qquad |\lambda|\ll 1,
	\label{bound0q}
\end{align}
is assumed to be small and may have either signs (the positive sign corresponds to repulsion, the negative -- to attraction). The dimensionless concentration $n$ is also assumed small. 

The Born decay rate
\begin{align}
	\frac{1}{\tau_{k,m}}=\frac{1}{\tau(E_{km})}=-2{\rm Im}\,\Sigma=\frac{1}{\tau_0}\frac{\nu(\varepsilon)}{\nu_0}.
	\label{life-time1}
\end{align}
For point impurities the scattering is isotropic and therefore the transport time coincides with the simple decay time.

 Thus, if \eqref{deni2} is fulfilled, the particle is scattered predominantly (though not completely) to the upper band. In particular, if the particle was already in the upper subband then the scattering event most probably will not remove it from there. It means that in the zero approximation the upper subband is almost  
decoupled from all others. 

Under the condition \eqref{deni2} the electrons in the $N$-subband states have low longitudinal velocity and therefore do not contribute much to the current. The latter is dominated by the states in all other bands. However,   the  singularity in the $N$-band is manifested also in the resistivity $\rho$ through the scattering rate that is proportional to the density of the final states on the Fermi surface:
 \begin{align}
	\frac{\rho}{\rho_0}=\frac{\nu(\varepsilon)}{\nu_0},\qquad	\rho_0=\frac{1}{e^2\varepsilon_0\tau_0}. \label{bound0qew}
\end{align}
These final states predominantly belong to the $N$-band. 
\subsection{Smearing of the Van Hove singularities within the Born approximation\label{Smearing of singularities within the Born approximation}}

It is clear that the scattering should somehow smear the singularities both in the density of states and in the resistivity. In this section we will discuss the mechanism of this smearing within the Born approximation. The condition of its applicability will be discussed in the following section. 

Within the Born approximation one can write (see, e.g., \cite{KearneyButcher1987, HuegleEgger2002})

\begin{align}
    	\langle\nu(\varepsilon)\rangle=\nu^{(0)}[\varepsilon-\Sigma(\varepsilon)]
	\label{selfff}
\end{align}
where $\Sigma	(\varepsilon)$ is given by \eqref{self-energy1}. We are interested in the behavior of $\langle\nu(\varepsilon)\rangle$ in the vicinity of the Van Hove singularity where $|\varepsilon|\ll 1$ and the scattering is dominated by the resonant band. It is convenient to discuss this problem separately for the cases $\varepsilon>0$ (above the singularity) and $\varepsilon<0$ (below the singularity).

\subsubsection{Above the Van Hove singularity\label{Above the Van Hove singularity}}

Here the self-energy is almost purely imaginary: the scattering is more important than the energy shift. The scattering obviously leads to decay of the plane waves and the decrement of this decay is just $\tau^{-1}$ given by the formula \eqref{life-time1}.
For $\tau^{-1}\ll\varepsilon$ the average density of states is almost insensitive to the scattering.

 In the narrow vicinity of the singularity, for $\tau^{-1}\sim\varepsilon$, the scattering becomes effectively strong, the density of states is strongly changing on the scale of the width of the relevant states. Thus, at $\tau^{-1}\sim\varepsilon$ the density of states saturates and we conclude that the corresponding peaks are smoothed  at $\varepsilon\sim\varepsilon_{\min}\sim\tau^{-1}$. However, $\tau^{-1}$ itself depends on $\varepsilon$ and so we arrive at the self-consistency condition:
\begin{align}
	\frac{1}{\tau(\varepsilon)}=\frac{1}{\tau_0}\frac{\nu(\varepsilon)}{\nu_0}\sim\frac{1}{\pi\tau_0\sqrt{\varepsilon}}\approx \varepsilon
	\label{self-consist1}
\end{align}
from where we can easily get
\begin{align}
	\varepsilon\sim\varepsilon_{\min}=(2\pi\tau_0)^{-2/3}=[(\lambda/\pi)^2(n/\pi)]^{2/3},\label{minieps01}
	\end{align}	
	\begin{align}
	 \frac{\tau_0}{\tau_{\min}}\sim\frac{\nu_{\max}}{\nu_0}\sim\frac{\rho_{\max}}{\rho_0}\sim\left[\left(\frac{\lambda}{\pi}\right)^2\left(\frac{n}{\pi}\right)\right]^{-1/3}.
	\label{minieps}
	\end{align}	
	\subsubsection{Below the Van Hove singularity\label{Below the Van Hove singularity}}
	
Now we have to find the density of states for $\varepsilon<0$. It seems clear that for 	$|\varepsilon|\lesssim \epsilon_{\min}$ the value of $\nu(\epsilon)$ can not change considerably so that one can expect 
\begin{align}
\nu(\varepsilon)\sim \nu_{\max},\quad \tau(\varepsilon)\sim \tau_{\min},\quad \mbox{for $|\varepsilon|\lesssim \varepsilon_{\min}$}
	\label{miniepsq}
	\end{align}
On the other hand, in the range $|\varepsilon|\gg \varepsilon_{\min}$ the correction to the density of states can be found with the help of perturbation theory which gives
\begin{align}
	\nu(\varepsilon)-\nu_0\sim\nu_0\frac{(-\varepsilon)^{-3/2}}{\tau_0}\sim\nu_0\left(\frac{\varepsilon_{\min}}{|\varepsilon|}\right)^{3/2}\ll\nu_0.
	\label{cor-den2}
	\end{align}
It is important to note that the correction \eqref{cor-den2} is relatively small already for $|\varepsilon|\lesssim\varepsilon_{\min}$. It means that
\begin{align}
	\nu(-\varepsilon_{\min})\sim\nu_0\ll \nu(\varepsilon_{\min})	\label{cor-den2lk}
	\end{align}
  and direct smooth matching of \eqref{cor-den2} and \eqref{miniepsq} is impossible! 
  
\begin{figure}[ht]
\includegraphics[width=0.9\linewidth]{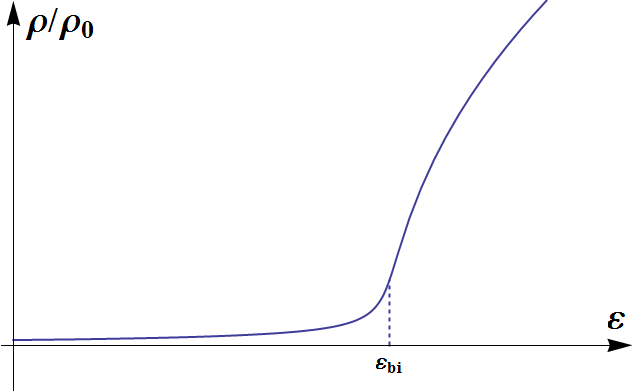}
\caption{The ``quasifold'': in the vicinity of the bifurcation point $\varepsilon=\varepsilon_{\rm bi}$ the slope of the curve $\rho(\varepsilon)$ is anomalously steep.}
\label{quasifold}
\end{figure}

  To resolve this paradox one should in principle go beyond the estimates made above, and accurately solve the problem in the range $|\varepsilon|\lesssim\varepsilon_{\min}$. However, for a qualitative understanding it is enough to note that there is practically only one scenario for such a giant drop in the density of states: a ``quasifold'' -- an inflection point with almost vertical slope, see Fig. \ref{quasifold}. 
  
    In the dependence $\nu(\varepsilon)$ at some point $\varepsilon_{\rm bi}$ there should be very large positive first derivative $\nu'(\varepsilon_{\rm bi})\gg\nu(\varepsilon_{\min})/\varepsilon_{\min}$ zero second derivative and rather small third derivative. An example of such a behaviour is provided by the results of the self-consistent Born approximation, given in Appendix \ref{self-consistent Born approximation}. Although these results can not be taken too seriously (since the self-consistent Born approximation is not rigorous), the main message seems to be reliable: the entire domain $|\varepsilon|\sim\varepsilon_{\min}$ is split into two basic subdomains: $\varepsilon<\varepsilon_{\rm bi}$ where $\nu\sim\nu_0$ and $\varepsilon>\varepsilon_{\rm bi}$ where $\nu\sim\nu(+\varepsilon_{\min})\gg \nu_0$. Between these two subdomains there is a narrow intermediate layer around  $\varepsilon_{\rm bi}$ in which $\nu(\varepsilon)$ undergoes a dramatic change.
  
\begin{figure}[ht]
\includegraphics[width=0.9\linewidth]{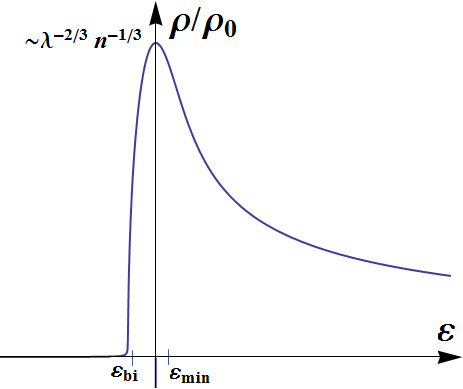}
\caption{The shape of a smeared Van Hove singularity within the Born approximation.}
\label{smeared-born}
\end{figure}

  Then the  results can be roughly summarized as follows:	
	\begin{align}
	\frac{\rho(\varepsilon)}{\rho_0}=\left\{
	\begin{aligned}
1+\frac{1}{\pi\sqrt{\varepsilon}},\quad  & \mbox{for $\varepsilon_{\min}\ll\varepsilon$, $\varepsilon>0$}, \\
\sim\lambda^{-2/3}n^{-1/3},\quad	&\mbox{for $\varepsilon_{\rm bi}<\varepsilon\lesssim\varepsilon_{\min}$}, \\
\sim 1,\quad	&\mbox{for $\varepsilon<\varepsilon_{\rm bi}$},
	\end{aligned}\right.
\label{self-consist2}		
\end{align}
with certain $\varepsilon_{\rm bi}<0$, $|\varepsilon_{\rm bi}|\sim \varepsilon_{\min}$. A schematic plot of \eqref{self-consist2} is shown in FIG.\ref{smeared-born}.

\section{Beyond the Born approximation\label{Beyond the Born approximation}}

The above considerations seem plausible and straightforward. However, the analysis below shows that they are only applicable if the concentration of impurities is high enough, i.e. for $n\gg n_c\sim|\lambda|$. For $n\ll n_c$ the scattering that determines the form of smeared Van Hove singularities is strongly modified by non-Born effects that dramatically grow upon approaching the singularity. We start our discussion from the properties of an exact amplitude of scattering by a single short-range impurity, placed on the wall of the tube.
 
\subsection{A single impurity problem in two dimensions: non-Born effects\label{A single impurity problem: non Born effects}}

Properties of short-range impurities or defects in two-dimensional systems  are  well studied. 
In this subsection we briefly remind the main facts. 

In particular, it is known that a weakly-attracting short-range impurity always forms a bound state \cite{opticaltheorem}. Writing the hamiltonian of the system in the form \eqref{imp-potential} with $\lambda<0$ one finds 
 that there is a single bound state with small binding energy
 \begin{align}
	E_{\rm bound}^{(2d)}\approx -\frac{\hbar^2}{m^*a^2_0}\exp\left(-\frac{\pi}{|\lambda|}\right),
	\label{bound0}
\end{align}
where $a_0$ is the ultraviolet cutoff (``radius of the delta-function'')
The wave-function of the ground state
  \begin{align}
	\psi_0(r)\sim\exp(-r/a^{(2d)}),\qquad a^{(2d)}=(2m^*|E_{\rm bound}^{(2d)}|)^{-1/2},
	\label{bound1}
\end{align}
being the radius of the  ground state wave function.

A scattering of a particle with positive energy $E\ll \frac{\hbar^2}{m^*a^2_0}$ is isotropic. For $r\gg a_0$ one can write the ``scattering wave-function'' in the form \cite{opticaltheorem}
 \begin{align}
	\psi_{\bf p}(r)=\exp\{i({\bf p\cdot r})\}-i\lambda H_0^{(1)}(pr),\quad E=p^2/2m^*,
	\label{bound1ss}
\end{align}
 where $H_0^{(1)}(x)$ is the Hankel function. Moreover, for $pr\gg 1$ one can write
  \begin{align}
	\psi_{\bf p}(r)\approx\exp\{i({\bf p\cdot r})\}-\lambda\sqrt{\frac{2}{-i\pi pr}}\exp(ipr),\qquad
	\label{bound1ss2}
\end{align}
 The above results should be modified if one wants to go beyond the Born approximation. If the condition $E\ll U_0$ (or $ka_0\ll 1$) is fulfilled then the scattering remains isotropic even beyond the Born approximation; it means that the scattering amplitude is still characterised by a single dimensionless  constant: small real $\lambda$ in the result \eqref{bound1ss2} should be replaced by not necessarily small complex $\Lambda$ -- the nonperturbative dimensionless scattering amplitude.  The latter should obey the optical theorem:
 \begin{align}
{\rm Im}\Lambda=-|\Lambda|^2,
	\label{unitarity10z}
\end{align}
hence the scattering amplitude can be parametrised by a single real constant $\lambda$
\begin{align}
\Lambda=\lambda e^{-i\arcsin\lambda}\equiv\lambda \left(\sqrt{1-\lambda ^2}-i\lambda \right).
	\label{unitarity20}
\end{align}
In particular, for weak interaction ($|\lambda|\ll 1$)
\begin{align}
\Lambda\approx\lambda-i\lambda^2.
	\label{unitarity20z}
\end{align}

Note that parameter $\lambda$ in \eqref{unitarity20} is related to the potential amplitude $V$ by formula \eqref{bound0q} only for $|\lambda|\ll 1$. In general case it is not true and  $\lambda$ is just a convenient parameter for expressing the phenomenological scattering amplitude.

\subsection{A single impurity problem on a cylinder: semiclassical treatment of non Born effects\label{semiclassical treatment of non-Born effects}}

 Let us place a single  weakly attracting impurity on the surface of the cylinder.  Clearly, there are two distinct cases with respect to the bound state of an electron:
 
(i) Wide cylinder or strong scattering: $R\gg a^{(2d)}$. In this case the bound state will not differ much from the purely two-dimensional case and the formula \eqref{bound0} applies.

(ii) Narrow cylinder or weak scattering: $R\ll a^{(2d)}$. This is an effectively one-dimensional case, the bound state can also be studied easily.

In this paper we will be interested, however, not in the ground state but in  the scattering  matrix for an electron with energy $E>0$ in the range
 \begin{align}
	E_0,E_{\rm bound}\ll E\ll \frac{\hbar^2}{m^*a^2_0}.
	\label{bounw1}
\end{align}
Under this condition the scattering process can be conveniently described in semiclassical terms. To find the scattering amplitude beyond the Born approximation one has to solve the Dyson equation
\begin{align}
G({\bf r_1,r_2})=	G_0({\bf r_1,r_2})+G_0({\bf r_1,r_0})VG({\bf r_0,r_2})
\label{dyson1}
\end{align}
for the retarded Green function defined as
  \begin{align}
	G=\left\{E-\hat{H}+i0\right\}^{-1}\quad G_0=\left\{E-\hat{H}_0+ i0\right\}^{-1}
	\label{imp-potential1}
\end{align}
where ${\bf r_0}$ is the position of the impurity. In particular, putting ${\bf r_1=r_2=r_0}$ we arrive at the equation
\begin{align}
g=	g_0+g_0Vg,\qquad g\equiv G({\bf r_0,r_0})=\frac{g_0}{1-Vg_0}
\label{dyson2}
\end{align}
where
\begin{align}
g_0\equiv G_0({\bf r_0,r_0})	
\label{dyson3}
\end{align}
One can also write
\begin{align}
G({\bf r_1,r_2})=	G_0({\bf r_1,r_2})+G_0({\bf r_1,r_0})V_{\rm ren}G_0({\bf r_0,r_2})
\label{dyson4}
\end{align}
with the renormalized scattering amplitude
\begin{align}
V_{\rm ren}=\frac{V}{1-Vg_0}
\label{dyson5}
\end{align}
First of all we have to find the single-site $g_0\equiv	G_{E}(0,0)$. For our nontrivial topology one can write in the semiclassical approximation
\begin{align}
g_0=\sum_{n=-\infty}^\infty e^{\pi in\Phi/\Phi_0}{\cal G}_{E}(2\pi nR),
	\label{green2}
\end{align}
${\cal G}_{E}({\bf r})$ being the retarded Green function in an infinite two-dimensional metal. For $n\neq 0$ one can use the semiclassical approximation: 
\begin{align}
{\cal G}_{E}({\bf r})\approx	\sqrt{\frac{2}{\pi pr}}e^{i( pr+\pi/4)}.
\label{green92}
\end{align}
For the $n=0$ term we have
\begin{align}
{\cal G}_{E}(0)=-\frac{im^*}{2}+C
	\label{green200}
\end{align}
where $C$ is a formally infinite real constant. This divergency is well known -- it means that the perturbation theory  does not work well in spatial dimensions $d\geq 2$ when applied to point-like impurities. This phenomenon is not specific for the cylinder geometry -- it is present in an infinite two-dimensional metal as well.
Special methods to deal with this divergence were developed already long ago. It was shown that in the case of isotropic scattering,  accurate calculations lead to the substitution of the bare coupling constant $\lambda$ by the exact complex  amplitude $\Lambda$ of scattering by the same impurity in the infinite two-dimensional metal. Thus, for the fully renormalized scattering amplitude $\Lambda^{\rm (ren)}$ in the case of cylinder we get
\begin{align}
\Lambda^{\rm (ren)}(\varepsilon)=\frac{\Lambda}{1+\Lambda g(\varepsilon)/\pi \nu_0},
	\label{unitarity3}
\end{align}
where $g(\varepsilon)\equiv g_0$ is given by the formulas \eqref{green2}, \eqref{green92}, \eqref{green200} where the infinite constant $C$ is discarded. As a result, we arrive at the expression \eqref{units}. Consequently, the scattering rate is also renormalized:
\begin{align}
\frac{1}{\tau(\varepsilon)}=\frac{2n}{\pi^2}\left|\frac{\Lambda}{1+\Lambda g(\varepsilon)/\pi \nu_0}\right|^2\frac{\nu(\varepsilon)}{\nu_0}.
	\label{unitarity3ssa}
\end{align}
For small $\lambda$ the discussed renormalization is only essential in the vicinity of some Van Hove singularity so that we can use asymptotics $g(\varepsilon)/\pi\nu_0\approx \pi^{-1}(-\varepsilon)^{-1/2}$ and for small $\lambda\ll 1$ one can write
\begin{align}
\Lambda^{\rm (ren)}(\varepsilon)\approx\frac{\lambda-i\lambda^2}{1+(\lambda-i\lambda^2)(-\varepsilon)^{-1/2}/\pi}.
	\label{unitarity3w}
\end{align}
The importance of the renormalization of the scattering matrix in the systems with the singularity in the density of states (e.g., superconductors) that can even lead to formation of bound states was discovered and explored in details already in 60-ies (see \cite{Fetter1965, MachidaShibata1972, Shiba1965, SodaMatsuuraNagaoka1967, Shiba1968}).

It is clear that the non-Born effects first come into play for $\varepsilon\lesssim \varepsilon_{\rm nB}$, where
\begin{align}
\varepsilon_{\rm nB}=(\lambda/\pi)^2,
	\label{unitarity3w43}
\end{align}
so that it is sometimes convenient to use the ``normalized'' energy:
\begin{align}
\epsilon\equiv \varepsilon/\varepsilon_{\rm nB}.
\label{self-consist2ssq}		
\end{align}

Note that for $\varepsilon\ll \varepsilon_{\rm nB}$ the scattering amplitude formally vanishes: $\Lambda^{\rm (ren)}\approx\pi(-\varepsilon)^{1/2}$. It means, in particular, that exactly at the van Hove singularity a quasi-one-dimensional system tends to become an ideal conductor with zero resistivity. In the following section we will demonstrate that for finite concentration of impurities the resistivity remains finite, though very small: it is proportional not to $n$, but to $n^3$. 

\section{ Single-impurity Non-Born effects in resistivity\label{Non-Born effects in resistivity}}

Physically the effect of renormalization is manifested in the scattering time $\tau(\varepsilon)$ in which $\lambda$ should be replaced by  $\Lambda^{\rm (ren)}(\varepsilon)$. Similar to the Born case, for 
\begin{align}
\tau^{-1}(\varepsilon)\ll \varepsilon
\label{self-consist2ssq22}		
\end{align}
the scattering is effectively weak (though non-Born!) so that only the single impurity effects should be taken into account and one can use the standard Drude formula with properly renormalized scattering time.  In this Section we concentrate on this ``weak non-Born scattering'' regime. We will consider the cases of repulsing and attracting impurities separately.

Certainly, there were some theoretical approaches  to the non-Born effects in quasi-one-dimensional systems in the past. S. H\"{u}gle and R. Egger \cite{HuegleEgger2002}  studied the smearing of  Van Hove singularities  within the self-consistent Born approximation similar to that described in Appendix \ref{self-consistent Born approximation}. In contrast with our work, instead of the quadratic spectrum of electrons they considered more realistic linear spectrum, characteristic of carbon nanotubes. This difference, however, is not essential, as far as one is interested only in the shape of the Van Hove singularities: it may actually be reduced to redefinition of some constants. What is much more important, instead of considering individual impurities S. H\"{u}gle and R. Egger introduced the disorder in the form of gaussian white noise. Such an approach does not allow to find the single-impurity non-Born effects that, as we have seen, are crucial at low concentrations $n_2\ll n_2^{(c)}$. So, their results are applicable to impurities only at high concentration $n_2\gg n_2^{(c)}$.

\subsection{Repulsing impurities}

For weak repulsive impurities ($\lambda>0$, $|\lambda|\ll 1$) the imaginary part of $\Lambda$ can be neglected and we get
\begin{align}
\frac{\rho(\varepsilon)}{\rho_0}=\frac{\tau_0}{\tau}=\frac{|\Lambda^{\rm (ren)}|^2}{\lambda^2}\left(1+\frac{1}{\pi\sqrt{\varepsilon}}\theta(\varepsilon)\right)=\nonumber\\=\left\{
	\begin{aligned}
\frac{1}{\lambda}\;
\frac{1}{\epsilon^{1/2}+\epsilon^{-1/2}},\quad  & \mbox{for $\epsilon>0$},\\
\frac{1}{(1+|\epsilon|^{-1/2})^2},\quad	&\mbox{for $\epsilon<0$},
	\end{aligned}\right.
\label{self-consist2ss}		
\end{align}
This dependence is plotted in Fig. \ref{resist-repuls}:

So, for $\varepsilon>0$ both the scattering rate and the resistivity have  smooth maxima at  $\varepsilon=\varepsilon_{\rm nB}$) with the value at maximum
\begin{align}
\frac{1}{\tau_{\min}^{(+)}}=\frac{1}{2\lambda\tau_0}=\frac{n\lambda}{\pi^2},\label{self-consist2ss09}		
\end{align}
or, in dimensional variables
\begin{align}
\frac{1}{\tau_{\min}^{(+)}}=\frac{2 n_2}{m^*}\lambda,\quad
\frac{\rho_{\max}^{(+)}}{\rho_0}=\frac{1}{\lambda}\gg 1.
\label{self-consist2ssm2o}		
\end{align}
For $\varepsilon<0$ the scattering rate grows monotonically with growing $|\varepsilon|$ and saturates at $\tau^{-1}=\tau_0^{-1}$ for   $|\varepsilon|\gg\varepsilon_{\rm nB}$ .

The non-Born effects somewhat suppress the resistivity, compared to the Born results. For repulsing impurities this is true for all $\varepsilon$ but the strongest effect is expected for $|\varepsilon|\lesssim \varepsilon_{\rm nB}$.

\subsection{Attracting impurities}

For attracting impurities the renormalized scattering amplitude has a pole in the complex plane of $\varepsilon$ at
\begin{align}
\varepsilon=\varepsilon_{\rm nB}(-1+2i\lambda),
\label{self-conser}		
\end{align}
close to the real axis. This fact indicates the existence of a quasistationary state.   We have to take into account the imaginary part of $\Lambda$ that keeps trace of the decay of this state: otherwise the pole would move to the real axis and there will be nonphysical divergence of amplitude. However, this is only necessary in the narrow vicinity of the resonance at $|\varepsilon|=\varepsilon_{\rm nB}$. So we can write
\begin{align}
\frac{\rho(\varepsilon)}{\rho_0}=\frac{\tau_0}{\tau}=\nonumber\\=\left\{
	\begin{aligned}
\frac{1}{|\lambda|}\;
\frac{1}{\epsilon^{1/2}+\epsilon^{-1/2}},\quad  & \mbox{for $\epsilon>0$},\\
\frac{1}{(1-|\epsilon|^{-1/2})^2},\quad	&\mbox{for $\epsilon<0$, $|1-|\epsilon||\gg |\lambda|$},\\
\frac{4}{(1-|\epsilon|)^2+4\lambda^2},\quad	&\mbox{for $\epsilon<0$, $|1-|\epsilon||\lesssim |\lambda|$},
	\end{aligned}\right.
\label{self-consist2ssk}		
\end{align}
This result is plotted Fig. \ref{resist-attract}.

Thus, for $\varepsilon>0$ (and also for $\varepsilon<0$ but $|\varepsilon|\ll \varepsilon_{\rm nB}$) the behaviour of the renormalized scattering rate for attracting impurities is identical to that of repulsing ones. Their behaviours are very different, however, for $\varepsilon<0$ (and not small $|\varepsilon|\gtrsim \varepsilon_{\rm nB}$). While for repulsive impurities both the rate $\tau^{-1}$ and the resistivity $\rho$ smoothly and monotonically increase with $|\varepsilon|$, for attracting impurities they first grow, reach sharp maxima at $\varepsilon=-\varepsilon_{\rm nB}$ and only then decrease, saturating at $\tau^{-1}=\tau_0^{-1}$ and $\rho=\rho_0$ for $|\varepsilon|\gg \varepsilon_{\rm nB}$. The maximum has a Lorenzian shape:
\begin{align}
\rho(\varepsilon)=\rho_{\max}^{(-)}\frac{\pi\Gamma_{\rm hom}}{2}L(\varepsilon+\varepsilon_{\rm nB},\,\Gamma_{\rm hom}),\label{self-consispppplu}\\ L(x,\gamma)\equiv\frac{\gamma/2}{\pi\left(x^2+(\gamma/2)^2\right)}.\label{self-consisppppl}		
\end{align}
The width of maximum (homogeneous broadening)
\begin{align}
\Gamma_{\rm hom}\sim 4|\lambda|\varepsilon_{\rm nB}=\frac{4|\lambda|^3}{\pi^2}\ll \varepsilon_{\rm nB},
\label{self-consispppp}
\end{align}
is relatively small. This decay is due to small probability of scattering to the bands other than the $N$-band. The height of the maximum is universal -- it does not depend on the strength of impurities $\lambda$. In dimensional variables:
\begin{align}
 \rho_{\max}^{(-)}=\frac{4n_2}{e^2m^*RE}.
\label{self-consist2ssm2}		
\end{align}
The scattering rate at maximum is even more universal:
\begin{align}
\frac{1}{\tau_{\min}^{(-)}}=\frac{1}{\lambda^2\tau_0}=\frac{2n}{\pi^2}=\frac{4n_2}{m^*}.
\label{self-consist2ssm1}		
\end{align}

\section{Multi-impurity effects. The central dip in resistivity.\label{Finite concentration of impurities}}

In the previous section we have implicitly assumed the concentration of impurities $n$ to be so low that scattering amplitude at certain impurity could not be affected by the presence of all the others: $\tau^{-1}(\varepsilon)\ll \varepsilon$. Let us first derive the condition that would justify this assumption. We have found that the non-Born effects are negligible for $\varepsilon\gtrsim \varepsilon_{\rm nB}$. On the other hand, if one totally neglects the non-Born effects, then, as it follows from \eqref{self-consist1}, the scattering effects lead to the saturation of both the density of states and the conductivity for $\varepsilon\lesssim \varepsilon_{\min}$. These two facts taken together mean that for $\varepsilon_{\rm nB}\ll \varepsilon_{\min}$ the non-Born effects do not have chance to show up at all. On the contrary, for $\varepsilon_{\min}\ll \varepsilon_{\rm nB}$ the scattering only comes into play at $\varepsilon\ll \varepsilon_{\rm nB}$ where the non-Born effects are already huge. Thus, looking at the expressions \eqref{minieps} for $\varepsilon_{\min}$ and \eqref{unitarity3w43} for $\varepsilon_{\rm nB}$ we conclude that the non-Born effects are relevant for $n<n_c$, where
\begin{align}
n_c\sim|\lambda|,
\label{self-consist2ssmw}		
\end{align}
while for $n>n_c$ the Born approximation is justified for all $\varepsilon$ and the results of section  \ref{Born scattering by short-range impurities} are applicable.

In this Section we are going to study the effect of scattering at low concentration $n\ll n_c$ but also at very low $|\varepsilon|$ at the same time. We will show that the presence of other impurities ultimately becomes essential in the narrow vicinity of the Van Hove singularity -- at certain energy scale $\varepsilon_{\min}^{\rm (nB)}$.

In the case of developed non-Born regime, for $\varepsilon\ll \varepsilon_{\rm nB}$ we have $\Lambda g\gg 1$, so that 
\begin{align}
\left|\Lambda^{\rm (ren)}(\varepsilon)\right|^2\approx\pi^2|\varepsilon|.
	\label{unitarity3wq}
\end{align}
We see that the rate $1/\tau$ ceases to depend on $\lambda$ and becomes universal: independent on the characteristics of impurities:
\begin{align}
	\tau^{-1}(\varepsilon)=2|\varepsilon|n\left(1+\frac{1}{\pi\sqrt{\varepsilon}}\theta(\varepsilon)\right).\label{life-timkd}
\end{align}
It should be stressed that the scattering rate decreases as the Fermi level approaches the Van Hove singularity from either side and formally vanishes at $\varepsilon=0$. Taken seriously, it would mean that exactly at singularity the system has zero residual resistivity. Of course, we expect that taking scattering in account will remove this paradox.

To demonstrate this, we have to incorporate the scattering in the result \eqref{life-timkd}. Again, as in Section \ref{Smearing of singularities within the Born approximation} we notice that the above calculations only make sense for $\tau^{-1}(\varepsilon)\ll \varepsilon$, so that the dip in the resistivity predicted by \eqref{life-timkd} will be rounded at certain $\varepsilon\sim \varepsilon_{\min}^{\rm (nB)}$, where $\varepsilon_{\min}^{\rm (nB)}$, however, is not given by \eqref{minieps01} any more because the expression for the scattering time \eqref{life-timkd} differs from \eqref{life-time1}: it has been changed by the non-Born effects. So, the self-consistency condition $\tau^{-1}(\varepsilon)\sim \varepsilon$ for $\varepsilon_{\min}^{\rm (nB)}$ reads
\begin{align}
\tau^{-1}\left(\varepsilon=+\varepsilon_{\min}^{\rm (nB)}\right)= \frac{2n}{\pi}\sqrt{\varepsilon_{\min}^{\rm (nB)}}\sim \varepsilon_{\min}^{\rm (nB)},
	\label{unitarity3wqi}
\end{align}
from where immediately follows
\begin{align}
\varepsilon_{\min}^{\rm (nB)}= \left(n/\pi\right)^2. 
	\label{unitarity3wqq}
\end{align}
Comparing \eqref{unitarity3wqq} to \eqref{unitarity3w43} we see that, indeed, the scattering effects bring the renormalization of the amplitude $\Lambda^{\rm (ren)}(\varepsilon)$ to stop at some small, but nonzero value.

The results \eqref{unitarity3wq} and  \eqref{unitarity3wqq} were obtained under the assumption $\varepsilon>0$ so we need yet to discuss the scattering effects for $\varepsilon<0$. Here we get
\begin{align}
\tau^{-1}(\varepsilon)=2n|\varepsilon|\ll |\varepsilon|,
	\label{unitarity3wo}
\end{align}
which formally means that for negative $\varepsilon$ the scattering does not affect the result \eqref{life-timkd} for all values of $|\varepsilon|$, down to $\varepsilon=0$! This is, of course, not quite true because, due to scattering effects, the discontinuity in the density of states at $\varepsilon=0$ should be smoothed and $1/\tau(\varepsilon)$ should remain of the order $1/\tau_{\max}$ also for $\varepsilon<0$ in the range $|\varepsilon|\lesssim \varepsilon_{\min}^{\rm (nB)}$. 

Thus, in the strongly non-Born domain $n\ll n^{\rm (nB)}$ we encounter the similar paradox as in the Born case at $n\gg n^{\rm (nB)}$. Namely, the above consideration  gives nonmatching estimates on the opposite sides of the interval $|\varepsilon|\lesssim\varepsilon_{\min}^{\rm (nB)}$:
\begin{align}
\tau^{-1}\sim\left\{\begin{aligned} n^2, &\quad\mbox{for $\varepsilon>0$, $\varepsilon\sim n^2$,}\\
n^3, &\quad\mbox{for $\varepsilon<0$, $|\varepsilon|\sim n^2$.}\end{aligned}\right.
	\label{unitarity3wobb}
\end{align}
The resolution of this paradox is also similar to that in the Born case: there is a quasifold at certain $\varepsilon=\varepsilon_{\rm bi}^{\rm (nB)}\equiv q_{\rm bi}\varepsilon_{\min}^{\rm (nB)}$, (with $q_{\rm bi}<0$, $|q_{\rm bi}|\sim 1$) where the scattering rate undergoes a dramatic drop, so that
\begin{align}
\tau^{-1}(\varepsilon)\sim\left\{\begin{aligned} 2n|\varepsilon|, &\quad\mbox{for $\varepsilon<\varepsilon_{\rm bi}^{\rm (nB)}$,}\\
n^2, &\quad\mbox{for $\varepsilon>\varepsilon_{\rm bi}^{\rm (nB)}$, $\varepsilon\lesssim \varepsilon_{\min}^{\rm (nB)}$,}\\
2n\sqrt{\varepsilon}/\pi, &\quad\mbox{for $\varepsilon\gg \varepsilon_{\min}^{\rm (nB)}$,}
\end{aligned}\right.
	\label{unitarity3wobby}
\end{align}
and the weakest scattering is realized at some $\varepsilon=\varepsilon_{\rm dip}^{\rm (nB)}$ below $\varepsilon_{\rm bi}^{\rm (nB)}$:
\begin{align}
\frac{1}{\tau_{\max}}\approx \frac{1}{\tau(\varepsilon=\varepsilon_{\rm dip}^{\rm (nB)})}\sim n\varepsilon_{\min}^{\rm (nB)}\sim n^3, 
	\label{unitarity3wqff}
\end{align}
or, in dimensional variables
\begin{align}
\frac{1}{\tau_{\max}}\sim\frac{(2\pi R)^4 [n_2]^3}{m^*}.
\label{self-consist2ssm2o}		
\end{align}

\begin{figure}[ht]
\includegraphics[width=0.9\linewidth]{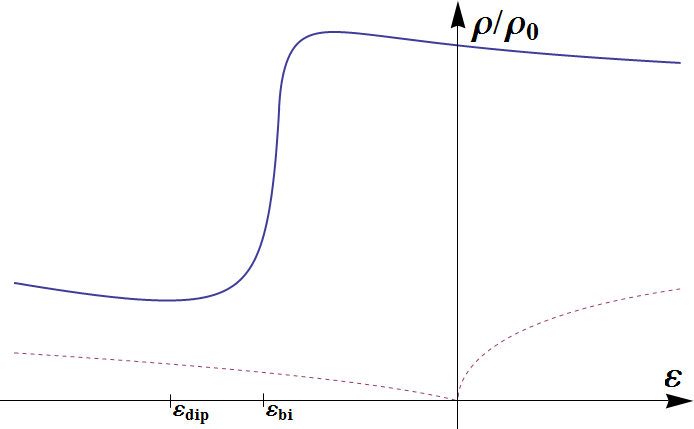}
\caption{The energy-dependence of resistivity near the minimum. Dashed line -- for $n\to 0$, solid line -- for finite $n$.}
\label{minimum}
\end{figure}

This result is supported by the calculations within the ``self-consistent non-Born approximation'', given in Appendix \ref{self-consistent non-Born approximation}.
Thus, we conclude that the minimal value of the scattering rate and, consequently, the minimal value of resistivity is attained a little bit to the left from the exact position of the Van Hove singularity, at $\varepsilon=\varepsilon_{\rm dip}^{\rm (nB)}\sim -n^2$ and 
\begin{align}
\rho_{\min}=\frac{1}{e^2RE}\frac{1}{\tau_{\max}}\sim \frac{(2\pi R)^4 [n_2]^3}{e^2m^*RE}
	\label{unitarity3wqs}
\end{align}
This minimal value depends neither on sign, nor on magnitude of $\lambda$ and is much less than the standard resistivity:
\begin{align}
\frac{\rho_{\min}}{\rho_0}\sim \frac{n^2}{\lambda^2}=\left(\frac{n}{n_c}\right)^2\ll 1.
	\label{unitarity3wju}
\end{align}
 The dependence $\rho(\epsilon)$ near the minimum is shown in Fig \ref{minimum}.
 
\section{Inhomogeneous contribution to  broadening of the resonant peak\label{Smearing of the resonant peak}}

One could expect that in the case of attracting impurities the scattering would lead also to broadening of the narrow resonant peak at $\varepsilon=-\varepsilon_{\rm nB}$, so that $\Gamma\to \Gamma_{\rm hom}+\tau^{-1}$. But this idea is wrong since the corresponding electrons are localized at resonant states  of certain individual impurities and, at low concentration, have no chance to be scattered by some other impurity. This statement is justified if $na_{\rm loc}\ll 1$, where $a_{\rm loc}=(2\varepsilon_{\rm nB})^{-1/2}=\pi|\lambda|^{-1}$ is the radius of the localized state. So, $na_{\rm loc}\sim n/|\lambda|\sim n/n_c$ and, under condition $n\ll n_c$,  the influence of other impurities {\it typically} is exponentially small. However, this influence may be large in some rare non-typical configurations and we will estimate their contribution.

Due to a rare local fluctuation two impurities may occur at non-typically small distance $r\lesssim a_{\rm loc}$ from each other, resulting in a considerable splitting $\Delta(r)\sim \varepsilon_{\rm nB}$ of a pair of initially degenerate localized states. It leads to inhomogeneous broadening
\begin{align}
\Gamma_{\rm inhom}\sim (na_{\rm loc})\varepsilon_{\rm nB}\sim\frac{n}{n_c}\varepsilon_{\rm nB}\label{self-consispi}		
\end{align}
that prevails in the intermediate range of concentrations: $|\lambda|^2\ll n\ll |\lambda|$, while for lowest $n\ll |\lambda|^2$ the homogeneous broadening is stronger. We should stress that the inhomogeneous broadening \eqref{self-consispi} exists already in the system where all impurities are identical (have the same $\lambda$). Naturally, the systems with dispersion of $\lambda$ demonstrate much stronger inhomogeneous broadening. We will briefly discuss such systems in Section \ref{Systems with different sorts of impurities}.

\section{Systems with different sorts of impurities\label{Systems with different sorts of impurities}}

In realistic physical systems the impurities are not necessarily identical. They may be of different types and they may be situated not directly in the wall of the tube, but at some distance from it. As a result the effective scattering amplitudes $\Lambda_i$ of different impurities may be different and random, with some distribution function $P(\lambda)$ for real parameter $\lambda_i$ (see \eqref{unitarity20}). The most important characteristic of this distribution is 
\begin{align}
\overline{\lambda}\equiv\sqrt{\langle\lambda^2\rangle}
\label{self-consisave2}	
\end{align}
What may be the consequences of such disorder?  In the Drude approximation the only dependence of the resistivity $\rho(\varepsilon)$ on $\Lambda$ comes from the factor $\tau^{-1}(\varepsilon)$. Since the contributions of different impurities to the resistivity are additive, one can write
\begin{align}
\langle\rho(\varepsilon)\rangle\propto\left\langle\frac{1}{\tau(\varepsilon,\lambda)}\right\rangle_{\lambda}\propto \int \left|\frac{\Lambda}{1+\Lambda g(\varepsilon)/\pi \nu_0}\right|^2P(\lambda)d\lambda\label{self-consisave}	
\end{align}
For $n\gg\overline{\lambda}$ the expression \eqref{self-consisave} can be expanded in small $\Lambda$, the non-Born effects are small  and we return to the results of Section \ref{Born scattering by short-range impurities} where one should substitute $\lambda\to\overline{\lambda}$.

For $n\ll\overline{\lambda}$ the scattering rate does not depend on $\lambda$ in the range $|\varepsilon|\ll \varepsilon_{\rm nB}\equiv(\overline{\lambda}/\pi)^2$, therefore all the results of Section \ref{Finite concentration of impurities} also apply to the case of random $\lambda$ in this range. The case $|\varepsilon|\sim \varepsilon_{\rm nB}$ is non-universal, here the result of averaging may depend on explicit shape of the function $P(\lambda)$. In particular, the contribution of the inhomogeneously broadened resonant peak can be evaluated with the help of expressions \eqref{self-consispppp}, \eqref{self-consispppplu}. Assuming that the Lorenzian peak in \eqref{self-consispppplu} is much sharper than the distribution $P(\lambda)$, we obtain
\begin{align}
\langle\rho^{\rm (res)}(\varepsilon)\rangle=\int d\lambda P(\lambda)\rho_{\max}^{(-)}\pi\Gamma_{\rm hom}(\lambda)\delta(\varepsilon+(\lambda/\pi)^2)=\nonumber\\=\rho_{\max}^{(-)}\pi^3|\varepsilon|P\left(-\pi\sqrt{|\varepsilon|}\right),\quad\mbox{for $\varepsilon<0$.}\label{self-consisavet}	
\end{align}

\section{Summary and discussion\label{discussion and conclusion}}

In this paper we have found the shape of the Van Hove singularity manifested in the resistivity of a clean metallic tube of radius $R$ with low concentration $n_{2}$ of weak short-range impurities (either repulsing or attracting) per unit surface of the tube. We have shown that there is certain crossover concentration $n_2^{(c)}=\frac{1}{(2\pi R)^2}\left(\frac{|\lambda|}{\pi}\right)^2$. For $n_2\gg n_2^{(c)}$ the  Van Hove singularities are smoothed peaks at $|E-E_N|\sim \Gamma_{\rm B}$ with the width
\begin{align}
    	\Gamma_{\rm B}\sim \frac{\hbar^2}{2m^*R^2}\left(\frac{R^2n_2|\lambda|^2}{\pi}\right)^{2/3}. \label{semic5yu}
\end{align}
The smoothing happens  due to interference of scattering events at different impurities, while the amplitude of scattering at each individual impurity is not affected. The structure of the Van Hove singularity for $n_2\gg n_2^{(c)}$ remains simple: ``plateau--maximum--plateau'' (see Fig. \ref{smeared-born}).

In the most interesting regime at $n_2\ll n_2^{(c)}$, the non-Born renormalization of individual scattering amplitudes happens already at $|E-E_N|\sim E_{\rm nB}$ where the interference effects are still negligible:
\begin{align}
    	E_{\rm nB}\sim \frac{\hbar^2}{2m^*R^2}\left(\frac{|\lambda|}{\pi}\right)^{2}\gg \Gamma_{\rm B}. \label{semic5yp}
\end{align}
Note that the energy scale $E_{\rm nB}$ does not depend on the concentration of impurities. The interference of scattering events at different impurities comes into play only at $|E-E_N|\sim \Gamma_{\rm nB}$, where the individual amplitudes are already strongly renormalized (suppressed) and take universal value
\begin{align}
   \lambda\to\lambda_0(E)=\left(\frac{2\pi^2 m^*R^2|E-E_N|}{\hbar^2}\right)^{1/2} \label{semic5yi}
\end{align}
which does not depend on the initial ``bare'' $\lambda$. As a result, instead of maximum $\rho(E)$ demonstrates a deep and narrow minimum at $E-E_N=E_{\rm dip}<0$ with a width
\begin{align}
    	\Gamma_{\rm nB}\sim|E_{\rm dip}| \sim \frac{\hbar^2}{2m^*R^2}\left(4\pi R^2n_2\right)^{2}\ll E_{\rm nB}. \label{semic5yl}
\end{align}
The structure of the Van Hove singularity for $n_2\ll n_2^{(c)}$ depends on the sign of the scattering amplitude: for repulsive interaction it is ``plateau--minimum--maximum--plateau'' (see Fig. \ref{resist-repuls}), while for attractive interaction it is ``plateau--maximum--minimum--maximum--plateau'' (see Fig. \ref{resist-attract}). We should stress, however, that an asymmetric structure of the Van Hove singularity, similar to the Fano resonance shape, arises at low concentration of impurities for both signs of the scattering amplitude.

In the leading approximation in small parameter $n_2/n_2^{(c)}$ (that corresponds to independent scattering at different impurities) the resistivity an minimum $\rho_{\min}$ vanishes, as it is shown in Figs. \ref{resist-repuls} and \ref{resist-attract}. The value of $\rho_{\min}$ becomes nonzero ($\rho_{\min}\propto n_2^3$, see Fig. \ref{minimum}) only  if one takes into account the interference of scattering events at different impurities.

In the future publications we are going to discuss the structure of Van Hove singularity in a conducting strip. Here the ``bare'' (non-renormalized) effective scattering amplitudes for different impurities inevitably differ from each other because of the random position of impuritiy with respect to the nodes of the transverse wave function in the resonant band. Since the dependence of the renormalized scattering amplitude on the bare one is non-monotonic, it can be shown that the leading contribution to the resistivity comes not from the ``strongest'' impurities (those sitting in the antinodes of the wave-function), but from some optimal ones. It leads to a serious modification of the results especially  in the range of small $|E-E_N|\ll E_{\rm nB}$.

In Conclusion, our study shows that at low concentration of impurities  the non-Born effects lead to  splitting of the Van Hove singularities in resistivity of a tube (or, in general, any other quasi-one-dimensional conductor) and this effect can not be described in terms of the Fano formula \eqref{fano}. The character of the splitting  depends on whether the impurities are attracting or repulsing.

We are indebted to M.V.Feigel'man, L.I.Glazman for illuminating discussions. 
This work was supported by Basic Research Program of The Higher School of Economics.

\appendix

\section{Self-consistent calculations: strong Born scattering\label{self-consistent Born approximation}}

Strictly speaking, the concept of the self energy is relevant only in the weak scattering domain where $|\varepsilon|\gg \varepsilon_{\min}$ (for both $\varepsilon>0$ and $\varepsilon<0$). However, using the perturbative expressions \eqref{selfff} and \eqref{self-energy1} also in the strong scattering domain $|\varepsilon|\ll \varepsilon_{\min}$ can be helpful for qualitative understanding of the behaviour of the density of states and resolving the paradox mentioned in the subsection \ref{Below the Van Hove singularity}. 

For a qualitative description of the density of states at strong scattering the self-consistent Born approximation can be used.
The selfconsistency equation for $\Sigma $ reads
\begin{align}
\Sigma(\varepsilon)=-\frac{i}{2\tau_0}\left(1+\frac{1}{\pi\sqrt{\varepsilon-\Sigma(\varepsilon)}}\right),
	\label{self-energy2}
\end{align}
so that
\begin{align}
\frac{\nu(\varepsilon)}{\nu_0}=1+{\rm Re}\,\left\{\frac{1}{\pi\sqrt{\varepsilon-\Sigma(\varepsilon)}}\right\},\label{self-energy3xd}\\
\Sigma(\varepsilon)=-\frac{i}{2\tau_0}-\varepsilon_{\min}Y\left[\frac{1}{\varepsilon_{\min}}\left(\varepsilon+\frac{i}{2\tau_0}\right)\right],
	\label{self-energy3}
\end{align}
where $Y(q)$ is the solution of cubic equation
\begin{align}
Y^2(Y+q)+1=0.
	\label{self-energy3a}
\end{align}
There is a bifurcation point $q=q_{\rm bi}$ such that for real $q<q_{\rm bi}$ all three solutions of \eqref{self-energy3a} are real while for $q>q_{\rm bi}$ there is one purely real solution and two conjugated  complex solutions (only the latter ones are physically relevant). Near the point $q=q_{\rm bi}$ one can write
\begin{align}
Y\approx Y_{\rm bi}\pm i A\sqrt{q-q_{\rm bi}}\\
Y_{\rm bi}=2^{1/3},\quad A=2^{2/3}3^{-1/2},\quad q_{\rm bi}=-3\cdot 2^{-2/3}.
	\label{self-energy4}
\end{align}
Thus, if the parameter $q$ were purely real then ${\rm Im}\,\Sigma$ would vanish for $\varepsilon< \varepsilon_{\rm bi}\equiv q_{\rm bi}\,\varepsilon_{\min}$. In our case, however, $q$ has small but finite imaginary part
\begin{align}
{\rm Im}\,q=\pi\sqrt{\varepsilon_{\min}}\ll 1.
	\label{self-energy3b}
\end{align}
For $\varepsilon> \varepsilon_{\rm bi}$ and $|\varepsilon- \varepsilon_{\rm bi}|\gg {\rm Im}\,q$ this imaginary part can be totally neglected and
\begin{align}
\Sigma(\varepsilon)\approx -\varepsilon_{\min}Y(\varepsilon/ \varepsilon_{\min}),\\
\frac{\nu(\varepsilon)}{\nu_0}=1+\frac{1}{\pi\sqrt{\varepsilon_{\min}}}\frac{1}{\sqrt{(\varepsilon/ \varepsilon_{\min})+Y(\varepsilon/ \varepsilon_{\min})}}.\label{einfach}
\end{align}
On the other side of the bifurcation point, for $\varepsilon< \varepsilon_{\rm bi}$ and $|\varepsilon- \varepsilon_{\rm bi}|\gg {\rm Im}\,q$ the ${\rm Im}\,q$-term may be taken into account perturbatively:
\begin{align}
\Sigma(\varepsilon)\approx-\varepsilon_{\min}Y\left(\varepsilon/\varepsilon_{\min}\right)-\frac{i}{2\tau_0}\left[1+Y'\left(\varepsilon/\varepsilon_{\min}\right)\right],\\
\frac{\nu(\varepsilon)}{\nu_0}=1+\frac{1+Y'\left(\varepsilon/\varepsilon_{\min}\right)}{2\left[(\varepsilon/\varepsilon_{\min})+Y(\varepsilon/ \varepsilon_{\min})\right]^{3/2}},
	\label{self-energy3we}
\end{align}
where $Y'(q)\equiv dY(q)/dq$.

In the narrow vicinity of the bifurcation point,
for  $|\varepsilon- \varepsilon_{\rm bi}|\lesssim {\rm Im}\,q$ one should keep ${\rm Im}\,q$ but, on the other hand, one can use expansion \eqref{self-energy4} for $Y(q)$. As a result, in this range we obtain
\begin{align}
{\rm Re}\,\Sigma(\varepsilon)\approx -\varepsilon_{\min}Y_{\rm bi},
\label{self-energy5w}
\end{align}
and
\begin{widetext}
\begin{align}
{\rm Im}\,\Sigma(\varepsilon)\approx -\frac{A}{2\tau_0}\left[\frac{\left(Q^2+1\right)^{1/2}+Q}{{\rm Im}\,q}\right]^{1/2}\approx-\frac{A}{2\tau_0\sqrt{{\rm Im}\,q}}\left\{\begin{aligned}\left(2 Q\right)^{1/2},&\quad Q>0,\;\;1\ll Q\ll ({\rm Im}\,q)^{-1},\\
\left(-2Q\right)^{-1/2},&\quad Q<0,\;\;1\ll |Q|\ll ({\rm Im}\,q)^{-1}.
\end{aligned}\right.
\label{longread1}
\end{align}
\begin{align}
\frac{\nu(\varepsilon)}{\nu_0}\approx\frac{1}{\pi\sqrt{\varepsilon_{\min}}}\frac{A\;{\rm Im}\,q}{2\left(q_{\rm bi}+Y_{\rm bi}\right)^{3/2}}\left[\frac{\left(Q^2+1\right)^{1/2}+Q}{{\rm Im}\,q}\right]^{1/2}\approx\nonumber\\\approx\frac{1}{\pi\sqrt{\varepsilon_{\min}}}\frac{A\sqrt{{\rm Im}\,q}}{2\left(q_{\rm bi}+Y_{\rm bi}\right)^{3/2}}\left\{\begin{aligned}\left(2Q\right)^{1/2},&\quad Q>0,\;\;1\ll Q\ll ({\rm Im}\,q)^{-1},\\
\left(-2Q\right)^{-1/2},&\quad Q<0,\;\;1\ll |Q|\ll ({\rm Im}\,q)^{-1}.
\end{aligned}\right.
	\label{self-energy3we}
\end{align}
\end{widetext}
Here
\begin{align}
 Q(\varepsilon)=2\tau_0\left(\varepsilon-\varepsilon_{\rm bi}\right).
\label{self-energy5}
\end{align}
So, as it is easy to check, for  $|Q|\gg 1$ the asymptotics \eqref{longread1} overlaps with the results \eqref{einfach} and \eqref{self-energy3we}.

The result \eqref{self-energy5} should not be taken too seriously: the self-consistency equation \eqref{self-energy3} can not be justified rigorously. However, the qualitative behaviour of the decay rate and the density of states predicted by  \eqref{longread1} and \eqref{self-energy3we}  gives us a reasonable pattern of matching conflicting results \eqref{minieps} and \eqref{cor-den2lk}. Namely, there is a narrow interval $|\varepsilon-\varepsilon_{\rm bi}|\lesssim 1/2\tau_0$ around certain  bifurcation point $\varepsilon_{\rm bi}$ ($\varepsilon_{\rm bi}<0$, $|\varepsilon_{\rm bi}|\sim\varepsilon_{\min}$) where both $\nu(\varepsilon)$ and $\tau^{-1}(\varepsilon)$ increase with $\varepsilon$ very rapidly, and just this increase explains the parametrically large difference between the results \eqref{minieps} and \eqref{cor-den2lk}.

\section{Self-consistent calculations: strong non-Born scattering\label{self-consistent non-Born approximation}}

The general (with an account for the non-Born renormalization of the scattering amplitude) self-consistency equation for the self-energy $\Sigma$ reads
\begin{align}
\Sigma(\varepsilon)=\frac{n}{\pi^2}\left|\Lambda^{\rm (ren)}(\varepsilon-\Sigma(\varepsilon))\right|^2\frac{g(\varepsilon-\Sigma(\varepsilon))}{\pi\nu_0}\
	\label{self-energy2xc0}
\end{align}
where $\Lambda^{\rm (ren)}$ is given by \eqref{unitarity3w} and the density of states is determined by formula \eqref{self-energy3xd}.
In the case of strong non-Born effect one can use an asymptotic expression \eqref{unitarity3wq} and get
\begin{align}
\Sigma(\varepsilon)=-in|\varepsilon-\Sigma|\left(1+\frac{1}{\pi\sqrt{\varepsilon-\Sigma(\varepsilon)}}\right),
	\label{self-energy2xc}
\end{align}
Let us first neglect the constant term in $g$, then we get
\begin{align}
\Sigma(\varepsilon)=-\frac{in}{\pi}\sqrt{\varepsilon-\Sigma^*},
	\label{self-energy2xc1}
\end{align}
or
\begin{align}
\Sigma=-\varepsilon_{\min}^{\rm (nB)}Y,\quad q\equiv\frac{\varepsilon}{\varepsilon_{\min}^{\rm (nB)}},\\
Y^2+q+Y^*=0.
	\label{self-energy2xc2}
\end{align}
For real $q$ the general structure of solutions for equation \eqref{self-energy2xc2} is as follows:

For $q>1/4$ there are two complex conjugated solutions:
\begin{align}
Y_{1,2}=\frac12\mp i\sqrt{\frac34 +q}
	\label{self-energy2xc3}
\end{align}
For $q<-3/4$ there are two real solutions:
\begin{align}
Y_{3,4}=-\frac12\mp\sqrt{\frac14 -q}
	\label{self-energy2xc4}
\end{align}
For $-3/4<q<1/4$ all four solutions $Y_{1,2,3,4}$ are acceptable.

There is, however, always only one physically relevant solution:
\begin{align}
Y(q)=\left\{\begin{aligned}Y_{4},\quad & \mbox{for $q<q_{\rm bi}$},\\
Y_{2},\quad & \mbox{for $q>q_{\rm bi}$}
\end{aligned}\right. \qquad q_{\rm bi}=-3/4.
	\label{self-energy2xc5}
\end{align}
 Thus, the bifurcation energy
 \begin{align}
\varepsilon_{\rm bi}^{\rm (nB)}=\varepsilon_{\min}^{\rm (nB)}q_{\rm bi},
	\label{self-energy2xc4q}
\end{align}
and for $\varepsilon_{\rm bi}^{\rm (nB)}<\varepsilon\ll\varepsilon_{\rm nB}$ we have
\begin{align}
\frac{1}{\tau(\varepsilon)}=-2{\rm Im}\,\Sigma(\varepsilon)=2\sqrt{\varepsilon_{\min}^{\rm (nB)}\left(\varepsilon-\varepsilon_{\rm bi}^{\rm (nB)}\right)},\label{self-energy3xdq}
\end{align}
\begin{align}
\frac{\nu(\varepsilon)}{\nu_0}=1+{\rm Re}\,\left\{\frac{1}{\pi\sqrt{\varepsilon+\varepsilon_{\min}^{\rm (nB)}Y_2\left(\varepsilon/\varepsilon_{\min}^{\rm (nB)}\right)}}\right\}=\nonumber\\=1+\frac{\sqrt{\varepsilon-\varepsilon_{\rm bi}^{\rm (nB)}}}{\pi\left(\varepsilon+\varepsilon_{\min}^{\rm (nB)}\right)},\label{self-energy3xdqa}
\end{align}
where we have used the formula
\begin{align}
\frac{1}{\sqrt{a+ib}}=\sqrt{\frac{\sqrt{a^2+b^2}+a}{2(a^2+b^2)}}-i\sqrt{\frac{\sqrt{a^2+b^2}-a}{2(a^2+b^2)}}.\label{self-eneerw}
\end{align}

So, the approximate equation \eqref{self-energy2xc1} leads to the result $1/\tau(\varepsilon)\equiv -2{\rm Im}\,\Sigma(\varepsilon)=0$ for all $\varepsilon<\varepsilon_{\rm bi}$. To determine finite scattering rate in this range we should go beyond and take into account the first term $-in|\varepsilon-\Sigma|$ on the right hand side of equation \eqref{self-energy2xc}. When doing so we can, however, substitute the found zero-approximation solution to this correction term. Then, instead of \eqref{self-energy2xc2}, we arrive at
\begin{align}
[Y+in(Y_4(q)+q)]^2+q+Y^*=0,
	\label{self-energy2xc6}
\end{align}
where we have noted that in the range $q<q_{\rm bi}$ both $q$ and $Y_4(q)$ are real, and also $q-Y_4(q)\equiv Y^2_4(q)$ is real and negative so that we could write $|\varepsilon-\Sigma|=\varepsilon_{\min}^{\rm (nB)}Y^2_4(q)$. Then
\begin{align}
\Sigma=-\varepsilon_{\min}^{\rm (nB)}[inY^2_4(q)+Y(\tilde{q})],
	\label{self-energy2xc7}
\end{align}
where $Y(\tilde{q})$ is the solution of \eqref{self-energy2xc2} with complex
\begin{align}
\tilde{q}\equiv q-inY^2_4(q).
	\label{self-energy2xc8}
\end{align}
For $q<q_{\rm bi}$ the imaginary part of $\tilde{q}$ can be treated perturbatively:
\begin{align}
Y(\tilde{q})\approx Y_4(q)+in\frac{Y^2_4(q)}{2Y_4(q)-1},
	\label{self-energy2xc9}
\end{align}
and
\begin{align}
\Sigma(q)\approx-\varepsilon_{\min}^{\rm (nB)}Y_4(q)\left\{1+in\frac{Y^2_4(q)}{Y_4(q)-1/2}\right\}
	\label{self-energy2xc10q}
\end{align}
\begin{widetext}
\begin{align}
\frac{1}{\tau(\varepsilon)}=-2{\rm Im}\,\Sigma(q)=
n\varepsilon_{\min}^{\rm (nB)}\frac{(\sqrt{1+4|q|}-1)^3(\sqrt{1+4|q|}+2)}{8(|q|-|q_{\rm bi}|)},\quad q\equiv \frac{\varepsilon}{\varepsilon_{\min}^{\rm (nB)}}<q_{\rm bi}\equiv-\frac34,
	\label{self-energy2xc10}\\
\frac{\nu(\varepsilon)}{\nu_0}=1+\frac{1}{2[Y_4(q)-1/2]}=1+\frac{\sqrt{\frac14+|q|}+1}{2(|q|-|q_{\rm bi}|)},
	\label{self-energy3we}
\end{align}
\end{widetext}
where we have used
\begin{align}
Y_4(q)-1/2=\sqrt{\frac14+|q|}-1=\frac{|q|-|q_{\rm bi}|}{\sqrt{\frac14+|q|}+1}.
	\label{self-energy2xyy}
\end{align}

In particular, for $|q|\gg1$ the scattering rate grows with $|\varepsilon|$ while $\nu(\varepsilon)$ saturates:
\begin{align}
\Sigma(\epsilon)\approx-\sqrt{\varepsilon_{\min}^{\rm (nB)}|\varepsilon|}-in|\varepsilon|,\quad \nu(\varepsilon)\approx\nu_0
	\label{self-energy2xc11}
\end{align}
which is in agreement with \eqref{unitarity3wo}. When $q$ approaches $q_{\rm bi}$ (i.e., $\varepsilon\to\varepsilon_{\rm bi}^{\rm (nB)}$ from below), both the scattering rate and the density of states grow:
\begin{align}
\Sigma\approx
-\frac{\varepsilon_{\min}^{\rm (nB)}}{2}\left\{1+\frac{in}{2}\frac{\varepsilon_{\min}^{\rm (nB)}}{\varepsilon_{\rm bi}^{\rm (nB)}-\varepsilon}\right\},\quad \frac{\nu(\varepsilon)}{\nu_0}\approx\frac{\varepsilon_{\min}^{\rm (nB)}}{\varepsilon_{\rm bi}^{\rm (nB)}-\varepsilon}.
	\label{self-energy2xc12}
\end{align}
So, the scattering rate reaches its minimum at some $\varepsilon=\varepsilon_{\rm dip}^{\rm (nB)}\equiv \varepsilon_{\min}^{\rm (nB)}q_{\rm dip}$, where 
\begin{align}
q_{\rm dip}=-\frac{21}{16},\quad \frac{1}{\tau(\varepsilon_{\rm dip})}=\frac{27}{8}n\varepsilon_{\min}^{\rm (nB)}	\label{self-energy2xc13}
\end{align}


\begin{thebibliography}{99}

\bibitem{VanHove} P.Y.Yu and M.Cardona {\it Fundamentals of Semiconductors.  Physics and Material Properties.} Chap. 6.2, Springer, (2010).
\bibitem{self-consistent} A. A. Abrikosov, L. P. Gorkov, and I. Dzyaloshinskii,
Quantum Field Theoretical Methods in Statistical Physics,
(Pergamon, New York, 1965).
\bibitem{self-consistent1} Elliott, R.I., Krumhansl, I.A., Leath, P.L.: Rev. Mod. Phys. {\bf 45}, 465 (1974).
\bibitem{self-consistent2} Lee, P.A., Ramakrishnan, T.V.: Rev. Mod. Phys. {\bf 57}, 287 (1985). 
\bibitem{Lee} P. A. Lee, Phys. Rev. Lett. {\bf 71}, 1887 (1993)
\bibitem{KearneyButcher1987} M.J. Kearney and P.N. Butcher, J. Phys. {\bf C20}, 47 (1987).
\bibitem{Peeters} P. Vasilopoulos,
F.M. Peeters, Revista Brasileira de Fysica, {\bf 19}, no 3, (1989)
\bibitem{HuegleEgger2002}S.~H\"{u}gle, R.~Egger, Phys. Rev. {\bf B 66}, 193311 (2002)



\bibitem{single-wall0}  Z. Zhang, D. A. Dikin, R. S. Ruoff, and V. Chandrasekhar, Europhysics Letters, {\bf 68}, 713 (2004).
\bibitem{single-wall}
B. Babi\'{c} and C. Sch\"{o}nenberger, Phys. Rev. {\bf B 70}, 195408 (2004)
\bibitem{multi-wall1}
 J. Kim, J. R. Kim, Jeong-O Lee, J. W. Park, H. M. So, N. Kim, K. Kang, K. H. Yoo, and J. J. Kim, Phys. Rev. Lett. {\bf 90}, 166403 (2003).
\bibitem{multi-wall2} W. Yi, L. Lu, H. Hu, Z. W. Pan, and S. S. Xie, Phys. Rev. Lett. {\bf 91}, 076801 (2003).
\bibitem{fano} U. Fano, Phys. Rev. {\bf 124}, 1866 (1961).
\bibitem{fano1}  A. E. Miroshnichenko, S. Flach, Y.S.Kivshar, Rev. Mod. Phys.  {\bf 82}, 2257 (2010)


\bibitem{Brandt77}  N.~B.~Brandt {\it et al},  Zh. Eksp. Teor. Fiz  {\bf 72}, 2332 (1977) [Sov. Phys. JETP, {\bf 45}, 1226 (1977)]; 
\bibitem{Brandt82}  N.~B.~Brandt {\it et al},  Fiz. Nizk. Temp.  {\bf 8}, 718 (1982) [Sov. J. Low Temp. Phys., {\bf 8}, 358 (1982)]
\bibitem{Nikolaeva2008}  A.~Nikolaeva {\it et al}, Phys. Rev. {\bf B 77}, 075332 (2008)

\bibitem{AB59}  Y.~Aharonov and D.~Bohm, Phys. Rev. {\bf 115}, 485 (1959)
\bibitem{AAS81}  B.~L.~Altshuler, A.~G.~Aronov, and B.~Z.~Spivak, Pis'ma Zh. Eksp. Teor. Fiz., {\bf 33}, 101 (1981) [JETP Letters, {\bf 33}, 94 (1981) 
\bibitem{AASSS82}  B.~L.~Altshuler, A.~G.~Aronov, B.~Z.~Spivak, D.~Yu.~Sharvin, and Yu.~V.~Sharvin, Pis'ma Zh. Eksp. Teor. Fiz., {\bf 35}, 476 (1982) [JETP Letters, {\bf 35}, 588 (1982)] 
\bibitem{AS87}  A.~G.~Aronov and Yu.~V.~Sharvin, Reviews of Modern Physics, {\bf 59}, 755 (1987) 
\bibitem{Ioselevich2015} A.~S.~Ioselevich, JETP Letters, {\bf 101}, 358 (2015)
\bibitem{spectrum} The quadratic spectrum is taken only for simplicity. Since the effects of interest are dominated by narrow vicinity of the Fermi level, the obtained results can be easily reformulated for arbitrary spectrum.
\bibitem{opticaltheorem} Landau and E.M.\, Lifshitz, Course in Theoretical
Physics (Pergamon, Oxford, 1981), Vol. 3 (Quantum mechanics. Nonrelativistic theory.)




















\bibitem{Fetter1965} A. L. Fetter,  Phys. Rev. {\bf 140},  A1921--A1936 (1965).
\bibitem{MachidaShibata1972} K. Machida and F. Shibata,  Prog. Theor. Phys. {\bf 47}, 1817 (1972).
\bibitem{Shiba1965} H. Shiba,  Prog. Theor. Phys. {\bf 50}, 50 (1973).
\bibitem{SodaMatsuuraNagaoka1967} T. Soda, T. Matsuura, and Y. Nagaoka,  Progress of Theoretical Physics {\bf 38}, 551 (1967).
\bibitem{Shiba1968} H. Shiba,  Progress of Theoretical Physics {\bf 40}, 435 (1968).







\end{thebibliography}
\end{document}